\documentclass[12pt,draftcls,onecolumn]{IEEEtran}

\usepackage{graphicx,amstext,amsmath,amssymb,psfrag,enumerate,color}
\usepackage{parskip,upgreek, times}
\usepackage{soul,here,stmaryrd}
\usepackage{epstopdf}
\usepackage{upgreek}
\usepackage{blindtext}
\usepackage{tikz}
\usepackage{array,multirow}
\newtheorem{theorem}{Theorem}
\newtheorem{lemma}[theorem]{Lemma}

\newtheorem{proposition}[theorem]{Proposition}

\newtheorem{assumption}{Assumption}
\newtheorem{remark}{Remark}

\def\0{{\bf 0}}
\def\x{{\bf X}}
\def\JJ{{\bf J}}
\def\DD{{\bf D}}
\def\AA{{\bf A}}
\def\R{\mathbb{R}}
\def\N{\mathbb{N}}

\def\I{{\bf I}}
\def\J{\mathcal{J}}

\def\II{\mathcal{I}}

\def\e{\mbox{e}}

\newcommand{\ie}{{\it i.e. }}

\usepackage{xcolor}
\usepackage{lipsum}
\newcommand{\blue}[1]{\textcolor{black}{#1}}

\graphicspath{{figures/}}

\begin{document}

\title{Stability analysis of a general class of\\ \blue{singularly perturbed linear hybrid systems} \thanks{This work was funded by the ANR project COMPACS - "Computation Aware Control Systems", ANR-13-BS03-004.} }

\author{Jihene Ben Rejeb, Irinel-Constantin Mor\u{a}rescu, Antoine Girard and Jamal Daafouz\thanks{J. Ben Rejeb, I.-C. Mor\u{a}rescu, J.Daafouz  are with Universit\'e de Lorraine, CRAN, UMR 7039 and CNRS, CRAN, UMR 7039, 2 Avenue de la For\^et de Haye, Vand\oe uvre-l\`es-Nancy, France.  e-mails: {\tt (jihene.ben-rejeb, constantin.morarescu, jamal.daafouz)@univ-lorraine.fr}.   A. Girard is with Laboratoire des signaux et syst\`emes (L2S), CNRS, CentraleSup\'elec, Universit\'e Paris-Sud, Universit\'e Paris-Saclay, 3, rue Joliot-Curie, 91192 Gif-sur-Yvette, cedex, France. e-mail:{\tt antoine.girard@l2s.centralesupelec.fr}}}

\maketitle

\begin{abstract}
 \blue{
Motivated by a real problem in steel production, we introduce and analyze a general class of singularly perturbed linear hybrid  systems with both switches and impulses, in which the slow or fast nature of the variables can be mode-dependent. This means that, at switching instants, some of the slow variables can become fast and vice-versa. 
Firstly, we show that using a mode-dependent variable reordering we can rewrite this class of systems in a form in which the variables preserve their nature over time. Secondly, we establish, through singular perturbation techniques, an upper bound on the minimum dwell-time ensuring the overall system's stability. Remarkably, this bound is the sum of two terms. The first term corresponds to an upper bound on the minimum dwell-time ensuring the stability of the reduced order linear hybrid system describing the slow dynamics. The order of magnitude of the second term is determined by that of the parameter defining the ratio between the two time-scales of the singularly perturbed system.
We show that the proposed framework can also take into account the change of dimension of the state vector at switching instants. Numerical illustrations complete our study.}
 \end{abstract}

\IEEEpeerreviewmaketitle

\section{Introduction}\label{sec:intro}

Systems characterized by processes that evolve on different time-scales are often encountered in biology  
\cite{ChenAihara2002,HodgkinHuxley} but are also present in engineering \cite{MallociPhD,SanfeliceTeel2011}. In this case, the standard stability analysis becomes more difficult and  singular perturbation theory \cite{KokotovicBook,KhalilBook} has to be used. This theory is based on Tikhonov approach that proposes to approximate the dynamics by decoupling the slow dynamical processes from the faster ones. The stability analysis is done separately for each time scale and under appropriate assumptions one can conclude on the stability of the overall system. Significant results related to stability analysis and approximation of solutions of  singularly perturbed systems can be found in \cite{Balachandra,NesicAveraging2001,TeelMoreauNesic}. 

Another feature that characterizes many physical systems is the presence of discrete events that occur during the continuous {evolution}. \blue{These events include
abrupt changes of dynamics or instantaneous state jumps, which lead to the classes of switched systems or impulsive systems, respectively.
Stability analysis and stabilization of singularly perturbed linear switched  systems are considered in \cite{Alwan-et-al2008,Malloci_et_al2010}. 
Interestingly, it is shown in~\cite{Malloci_et_al2010} that even though the switched dynamics on each time scale are stable, the overall system may be destabilized by fast switching signals. Clearly, this is in contrast with classical results on continuous singularly perturbed linear systems~\cite{KokotovicBook} and is a motivation for developing dedicated techniques for stability analysis of singularly perturbed hybrid systems. Stability analysis of singularly perturbed impulsive systems is considered in~\cite{Simeonov1988,AbdelrahimPostoyan2015}.
More general  singularly perturbed hybrid systems can involve both switches and impulses. A stability  result for this class of systems can be found in \cite{SanfeliceTeel2011}.
In these works, the slow or fast nature of the state variable does not change when an event (switch or impulse) occurs. In this paper we introduce and analyze a class of singularly perturbed linear hybrid systems in which, at switching instants, slow variables can become fast and vice-versa. Our framework also includes the analysis of singularly perturbed linear systems with or without switches and/or impulses.} \blue{ Moreover, taking advantage of the linear dynamics under study, we go beyond  the results in \cite{SanfeliceTeel2011} by characterizing the required dwell-time in terms of the parameter defining the ratio between the two time-scales.}

The analysis of the class of dynamical systems discussed in this paper is motivated by an industrial application. In steel production, steering control denotes strategies to guide a metal strip in a finishing mill, which is constituted by a fixed number of stands. Each stand contains a set of rolls that crush the strip. The objective in rolling mills is to reduce the thickness of a strip. This goal is reached by maintaining the strip in a straight line and close to the mill axis, avoiding sudden lateral movements of the rolled product. As long as the strip remains connected to the coilbox, which is the device used to coil the strips into the finishing train, the hot strip model is described by a set of classical non linear differential equations. Indeed, each stand is linked to the others by the strip traction and there is no discontinuity in the model. The corresponding control problem can be treated using classical linear techniques motivated by the fact that it is enough from a practical point of view to consider small deviations around an ideal operating point (see \cite{Mallocietal2010} and references therein). One has to take care of the two time scale nature of this system as there is a slow dynamics corresponding to the lateral displacement of the strip after each stand (called strip off-centre) and a fast dynamics corresponding to the angle between the strip and the mill axis. As explained in \cite{Mallocietal2010}, using singular perturbations and time-scale separation, it is possible to design an efficient robust control strategy based on the reduced model which describes only the slow dynamics. Such a control law has been validated on the industrial plant.

The situation is different in the last phase of the rolling process called the tail end phase and where the strip leaves the stands one after the other. Traction is lost each time the strip leaves a stand and this increases the difficulty to guide the strip as it is free to move in all directions. There are several difficulties in this phase. The first one is related to model discontinuities. Each time the strip leaves a stand the system dynamics changes and switching occurs. Moreover, the tail end phase is very short, the switchings are very fast and stability of all subsystems is not a sufficient condition to guarantee the stability of the whole system. Without stability guarantee, these switchings may lead to a crash damaging the rolls. The second difficulty is related to the changes in the nature of the dynamics after switching. When the strip leaves the coilbox, the slow dynamics is given by the strips off-centre of the operating stands and the angle corresponding to the first active stand where the traction is lost. The angle which was a fast variable before leaving the coilbox becomes a slow variable in the tail end phase. This change occurs at each time the strip leaves a stand which means that the components and the dimension of the state vector change at each switching time.  A system with this behaviour can be defined as a switched system with multiple time scales, changes in the nature of the dynamics associated to each state variable and changes in the dimension of the state vector \cite{Mallocietal2009}. 

\blue{
Starting from the above motivation, we analyze a general class of singularly perturbed linear hybrid systems with mode-dependent nature of the state variable. 
The main contributions of the current work are:
\begin{itemize}
\item a procedure to rewrite the general class under study as linear hybrid singularly perturbed systems where the nature of variables does not change at switching instants, both cases of fixed and variable dimensions of the slow and fast state vectors are considered;
\item a new approach for stability analysis of singularly perturbed linear hybrid systems with both switches and impulses;
\item the derivation of an  upper bound on the minimal dwell-time between two events that ensures the stability of the singularly perturbed linear hybrid system.
\end{itemize}
It is noteworthy that, this bound is given as the sum of two terms. 
The first one corresponds to an upper bound on the minimum dwell-time ensuring the stability of the reduced order linear hybrid system describing the slow dynamics. The order of magnitude of the second term is determined by that of the parameter $\varepsilon$ defining the ratio between the two time-scales of the singularly perturbed system.
In particular, it follows that when the reduced order system has a common quadratic Lyapunov function, the first term is zero and the minimum dwell-time ensuring the stability of the overall system goes to zero as fast as $\varepsilon$ or $-\varepsilon \ln (\varepsilon)$
when the time scale parameter $\varepsilon$ goes to zero.}

Basically, we combine the classical singular perturbation theory \cite{KokotovicBook}  with Lyapunov function arguments for hybrid systems (see \cite{Teel-book} for details). 
\blue{Our results clearly differ from existing ones on  singularly perturbed  linear hybrid systems that we mentioned previously:
\cite{Malloci_et_al2010} deals with the existence of common quadratic Lyapunov functions and thus characterizes systems that are stable without dwell-time assumption;
the condition on the dwell-time established in~\cite{Alwan-et-al2008} does not present a clear separation between the slow and fast dynamics of the system;
and in~\cite{Simeonov1988,AbdelrahimPostoyan2015,SanfeliceTeel2011} the stability is established under a dwell-time condition where the dwell-time does not explicitly depend on the time-scale parameter.}

{The paper is organized as follows : Section \ref{Problem formulation} describes the hybrid system model in the singular perturbation form and introduces the relevant notations. In this section, we also introduce a mode-dependent reordering of the state components allowing to rewrite the system in a form in which the variables preserve their nature over time. Section \ref{Preliminaries} is devoted to preliminary results concerning the stability analysis of singularly perturbed linear systems without switches or jumps. Section \ref{Sec:stability} presents the main results along with their Lyapunov-based proofs. These results give stability conditions and establish an upper-bound on the minimum dwell-time ensuring the stability of the system. An extension to the case of mode-dependent dimension of the state-vector is provided in Section \ref{extension}. To illustrate the results,  we provide in Section \ref{sec:scalar} a dwell-time analysis and a numerical example in the particular case of scalar fast and slow dynamics with only two switching modes. Some concluding remarks end the paper.}

\section*{Notation}
Throughout this paper, $\R_+$ , $\R^n$ and $\R^{n\times m}$ denote respectively, the set of nonnegative real numbers, the $n$ dimensional Euclidean space and the set of all $n \times m$ real matrices.
The identity matrix of dimension $n$ is denoted by $\I_n$. We also denote by $\0_{n,m} \in \R^{n \times m}$ the matrix whose components are all $0$.
For a matrix $A \in \R^{n \times n}$, $\|A\|$ denotes the spectral norm \ie induced 2 norm. $A \ge \0$ $(A\le  \0)$ means that $A$ is positive semidefinite (negative semidefinite).
We write $A^{\top}$ and $A^{-1}$ to respectively denote the transpose and the inverse of $A$. For a symmetric matrix $A \ge \0$, $A^{\frac{1}{2}}$ is the unique symmetric matrix 
$B \ge 0$ such that $B^2=A$.
The matrix $A$ is said to be Hurwitz if all its eigenvalues have negative real parts. $A$ is said to be Schur if all its eigenvalues have modulus smaller than one. The matrix $A$ is said to be positive if all its coefficients are positive.
We also use $x(t^-)=\displaystyle\lim_{\delta\rightarrow 0,\ \delta>0} x(t - \delta)$.
Given a function $\eta:(0,\varepsilon^*)\rightarrow \R$, we say that $\eta(\varepsilon)=\mathcal O(\varepsilon)$ if and only if there exists $\varepsilon_0 \in (0,\varepsilon^*)$ and
$c>0$, such that for all $\varepsilon \in (0,\varepsilon_0)$, $|\eta(\varepsilon)| \le c \varepsilon$.

%

\section{Problem formulation}\label{Problem formulation}

In this paper, we consider a general class of singularly perturbed linear hybrid (i.e. switched and impulsive) systems. This class encompasses the case in which some slow varying variables switch to fast variation and/or reversely fast varying variables switch to slow variation. 

In order to formalize the system dynamics, let $\varepsilon >0$ be the small parameter characterizing the time scale separation between the slow and the fast dynamics. We consider a finite set of indices $\II$ and we introduce the diagonal matrices $\DD^i$ for all $i\in \II$. Precisely, the diagonal elements of each $\DD^i,\ i\in\II$ belong to the set $\{\varepsilon,1\}$ and they are used to select the fast and slow variables as explained below. 
We study switched systems of the form:
\begin{equation}\label{general_fast_slow_dyn}
 \DD^{\sigma_k}\dot{\x}(t)=\AA^{\sigma_k}\x(t), \ \forall t\in [t_k,t_{k+1}), \; k\in \N
\end{equation}
with impulsive dynamics :
\begin{equation}\label{original_reset-map}
\x(t_k)=\JJ^{\nu_k}\x(t_k^-), \quad \forall k \ge 1
\end{equation}
where $\x(t)\in\R^{n_{\sigma_k}},\ \forall t\in [t_k,t_{k+1}), \; k\in \N$,  and $0=t_0 < t_1 < \dots$  \blue{is the monotonically increasing and unbounded sequence of instants} of discrete events (switches or impulses), $\sigma_k\in \II$ and $\nu_k \in \J$ with $\II$ and $\J$ finite sets of indices.  
For all $k\in \N$,
$\AA^{\sigma_k},\DD^{\sigma_k}\in\R^{n_{\sigma_k}\times n_{\sigma_k}}$ and $\JJ^{\nu_k}\in\R^{n_{\sigma_k}\times n_{\sigma_{k-1}}}$ are matrices defining the continuous and impulsive dynamics. 

For all $i\in\II$, the matrix $\DD^i$ is used to specify the slow and fast varying variables as follows: 
\begin{itemize}
\item the $h$-th component of $\x$ has a fast variation when $\sigma_k=i$ if the $h$-th diagonal element of $\DD^i$ equals $\varepsilon$;
\item the $h$-th component of $\x$ has a slow variation when $\sigma_k=i$ if the $h$-th diagonal element of $\DD^i$ equals $1$.
\end{itemize} 
In the sequel, we will mainly focus on the case where the dimension of $\x$ is time-invariant (\ie $n_{i}=n,\ \forall i\in\II$) and the number of slow and fast varying variables remains constant. In other words, the number of entries of $\DD^i$ equal to $\varepsilon$ is constant, denoted by $n_z\le n$ for all $i\in \II$. This means that $\x$ has $n_z$ fast varying components and $n_x=n-n_z$ slow varying ones. 
\blue{This is without loss of generality, as we shall see in Section \ref{extension} that the case of time-varying dimensions $n$, $n_z$ and $n_x$ can be reduced to the case of fixed dimensions  by adding artificial stable variables.}

\begin{remark}
The stability analysis of \eqref{general_fast_slow_dyn}-\eqref{original_reset-map} encompasses the analysis of several existing classes of singularly perturbed linear hybrid systems. To illustrate that, let us suppose that $\DD^i=\DD^j, \forall i,j\in\II$ and denote by $x$ and $z$ the vectors of slow and fast components of $\x$, respectively. Then, system \eqref{general_fast_slow_dyn}-\eqref{original_reset-map} becomes a singularly perturbed linear hybrid system of the form:
\[
\left\{\begin{split}
 \left(\begin{array}{c}
\dot {x}(t)\\
\varepsilon \dot {z}(t)\\
\end{array}\right)
&=\AA^{\sigma_k} \left(\begin{array}{c}
x(t)\\
z(t)
\end{array}\right), \
 \ \forall t\in [t_k,t_{k+1}), \; k\in \N \\
\left(\begin{array}{c}
x(t_k)\\
z(t_k)
\end{array}\right)&=\JJ^{\nu_k} \left(\begin{array}{c}
x(t_k^-)\\
z(t_k^-)
\end{array}\right)\end{split}\right.
\]  
\blue{We can then trivially recover singularly perturbed switched systems (when there is only one jump matrix given by the identity) and singularly perturbed impulsive systems (when there is only one flow matrix), which are studied in~\cite{Alwan-et-al2008,Malloci_et_al2010} and in~\cite{Simeonov1988,AbdelrahimPostoyan2015}, respectively. We also point out that this class of systems is a subclass of singularly perturbed hybrid systems studied in \cite{SanfeliceTeel2011}.
Fundamental differences between our approach and these works have been highlighted in the introduction.}

\end{remark}

\subsection{Variable reordering}

A first step in our analysis is to rewrite \eqref{general_fast_slow_dyn} in a form where slow/fast variables remain slow/fast over time, independently of switches affecting the system's dynamics. To accomplish this step,  for all $i\in\II$ we introduce the permutation matrix $S_i$ such that 
\begin{equation}\label{D-transf}
S_i\DD^i S_i^\top=\left(\begin{array}{cc} \I_{n_x} & \0_{n_x,n_z}\\
\0_{n_z,n_x} &  \varepsilon\I_{n_z}\end{array}\right),\quad \forall i\in\II
\end{equation} and define the time dependent change of variable
\begin{equation}\label{var-transf}
\left(\begin{array}{c}
x(t)\\
z(t)
\end{array}\right)=S_{\sigma_k}\x(t), \ \forall t\in [t_k,t_{k+1}), \; k\in \N
\end{equation}
where $x(t)\in \R^{n_x}$,
$z(t)\in \R^{n_z}$.
In other words, we use the matrix $S_i$ to permute the components of $\x$ such that the first $n_x$ ones are characterized by a slow variation while the rest of $n_z$ components have a fast variation. Let us also introduce the following matrices:
\blue{
\begin{equation}\label{dyn-transf}
A^{i}= S_{i}\AA^{i} S_{i}^\top, \ J^{i\stackrel{j}{\rightarrow} i'}= S_{i'}\JJ^{j}S_{i}^\top,\; \forall i,i' \in\II,\ j \in\J.
\end{equation}
}
Using the change of variable \eqref{var-transf} and taking into account the matrices definitions \eqref{D-transf} and \eqref{dyn-transf}, the general system \eqref{general_fast_slow_dyn}-\eqref{original_reset-map} is rewritten in the following equivalent form:
\begin{equation}\label{equation_original}
\left(\begin{array}{c}
\dot {x}(t)\\
\varepsilon \dot {z}(t)\\
\end{array}\right)
=A^{\sigma_k}
\left(\begin{array}{c}
x(t)\\
z(t)
\end{array}\right), \ \forall t\in [t_k,t_{k+1}), \; k\in \N
\end{equation}
with impulsive dynamics:
\blue{
\begin{equation}\label{reset-map}
\left(\begin{array}{c}
x(t_k)\\
z(t_k)
\end{array}\right)=J^{\sigma_{k-1}\stackrel{\nu_{k}}{\rightarrow} \sigma_{k}} \left(\begin{array}{c}
x(t_k^-)\\
z(t_k^-)
\end{array}\right), \quad \forall k \ge 1
\end{equation}
}
%
\begin{remark} Switches and impulses can, but need not, be concomitant. Indeed, \blue{if $\sigma_k=\sigma_{k-1}$ and $J^{\sigma_{k-1}\stackrel{\nu_{k}}{\rightarrow} \sigma_{k}}\ne \I_{n}$}, then at time $t_k$ an impulse occurs but no switch.
Similarly, \blue{if $\I_{n} \in \{J^{i\stackrel{j}{\rightarrow} i'} |\; i,i'\in \II, j\in \J\}$, then if $J^{\sigma_{k-1}\stackrel{\nu_{k}}{\rightarrow} \sigma_{k}}=\I_{n}$ and $\sigma_k\ne \sigma_{k-1}$}, then at time $t_k$ a switch occurs but no impulse. 
\end{remark}

In general, stability analysis of \eqref{equation_original}-\eqref{reset-map} is a difficult task as it cannot be reduced to the analysis of the associated reduced (slow) and boundary layer (fast) systems, \blue{as shown in~\cite{Malloci_et_al2010}}. \blue{In the following, we will provide a new methodology based on singular perturbation techniques to characterize an upper-bound on the minimum dwell-time ensuring stability}. 


\subsection{Change of variable}

\blue{
For $i,i'\in \II$, $j\in \J$, let 
\begin{equation*}
A^i
=\left(\begin{array}{cc}A_{11}^{i} & A_{12}^{i}\\
A_{21}^{i} & A_{22}^{i}\end{array}\right),\;
J^{i\stackrel{j}{\rightarrow} i'}
=\left(\begin{array}{cc}J_{11}^{i\stackrel{j}{\rightarrow} i'} & J_{12}^{i\stackrel{j}{\rightarrow} i'}\\
J_{21}^{i\stackrel{j}{\rightarrow} i'} & J_{22}^{i\stackrel{j}{\rightarrow} i'}\end{array}\right),
\end{equation*}
where $A_{11}^{i}, J_{11}^{i\stackrel{j}{\rightarrow} i'} \in \R^{n_x\times n_x}$, and $A_{22}^{i}$, $A_{12}^i$, $A_{21}^i$, $J_{22}^{i\stackrel{j}{\rightarrow} i'}$, $J_{12}^{i\stackrel{j}{\rightarrow} i'}$, $J_{21}^{i\stackrel{j}{\rightarrow} i'}$ are of appropriate dimensions.
}

Let us impose the following standard assumption \cite{KokotovicBook} in the singular perturbation theory framework:
\begin{assumption}\label{assumption_A22}
$A_{22}^{i}$ is non-singular for all $i \in \II$.
\end{assumption}

Then, we perform the following time dependent change of variable:
\begin{equation}\label{y_change}
\left(\begin{array}{c}x(t)\\ y(t)\end{array}\right)=P_{\sigma_k}\left(\begin{array}{c}x(t)\\ z(t)\end{array}\right) ,\; \forall t\in [t_k,t_{k+1}), \; k\in \N
\end{equation}
where, for all $i\in \II$
\begin{equation*}
P_{i}=\left(\begin{array}{cc}\I_{n_x} & \0_{n_x,n_z}\\
(A_{22}^{i})^{-1}A_{21}^{i}&  \I_{n_z}\end{array}\right).
\end{equation*}
It is worth noting that the matrix $P_i$ is invertible and for all $i\in \II$
\begin{equation*} P_{i}^{-1}=\left(\begin{array}{cc}\I_{n_x} & \0_{n_x,n_z}\\
-(A_{22}^{i})^{-1}A_{21}^{i}&  \I_{n_z}\end{array}\right).
\end{equation*}

Using \eqref{y_change}, the continuous dynamics \eqref{equation_original} in the variables $x,y$ becomes:
\begin{equation}\label{system_fast_slow}
\begin{split}
 \left(\begin{array}{c}
\dot {x}(t)\\
\varepsilon \dot {y}(t)\\
\end{array}\right)
=\left(\begin{array}{cc}A_0^{\sigma_k} & B_1^{\sigma_k}\\
\varepsilon  B_2^{\sigma_k} & A_{22}^{\sigma_k}+\varepsilon B_3^{\sigma_k}\end{array}\right)\left(\begin{array}{c}
x(t)\\
y(t)
\end{array}\right), \\ 
 \ \forall t\in [t_k,t_{k+1}), \; k\in \N \end{split}
\end{equation}
where for all $i\in\II$ one has
\begin{equation*}
\begin{split}
& A_{0}^{i} = A_{11}^{i}-A_{12}^{i} (A_{22}^{i})^{-1} A_{21}^{i}, \ B_{1}^{i}= A_{12}^{i}, \\ & B_2^{i}=(A_{22}^{i})^{-1} A_{21}^{i} A_{0}^{i}, \ B_3^{i}= (A_{22}^{i})^{-1} A_{21}^{i} A_{12}^{i}.
\end{split}
\end{equation*}
Similarly, the impulsive dynamics \eqref{reset-map} is rewritten in the $x,y$ variables as:
\begin{equation}\label{reset-map_fast-slow}
\left(\begin{array}{c}
x(t_k)\\
y(t_k)
\end{array}\right)=R^{\sigma_{k-1}\stackrel{\nu_k}{\rightarrow} \sigma_{k}} \left(\begin{array}{c}
x(t_k^-)\\
y(t_k^-)
\end{array}\right), \quad \forall k \ge 1 \\
\end{equation}
where for all $i,i' \in \II$, $j\in \J$,
\blue{
\begin{equation*}
R^{i\stackrel{j}{\rightarrow} i'}=P_{i'}J^{i\stackrel{j}{\rightarrow} i'} P_{i}^{-1} = 
\left(\begin{array}{cc} R^{i\stackrel{j}{\rightarrow} i'}_{11} & R^{i\stackrel{j}{\rightarrow} i'}_{12}\\
R^{i\stackrel{j}{\rightarrow} i'}_{21} &R^{i\stackrel{j}{\rightarrow} i'}_{22} \end{array}\right)
\end{equation*}
}
with
\blue{
\begin{equation*}
\begin{split}
R^{i\stackrel{j}{\rightarrow} i'}_{11} =\; & J_{11}^{i\stackrel{j}{\rightarrow} i'} -J_{12}^{i\stackrel{j}{\rightarrow} i'}(A_{22}^{i})^{-1}A_{21}^i, \\
R^{i\stackrel{j}{\rightarrow} i'}_{12} =\; &  J_{12}^{i\stackrel{j}{\rightarrow} i'}, \\
 R^{i\stackrel{j}{\rightarrow} i'}_{21} =\; & (A_{22}^{i'})^{-1}A_{21}^{i'}(J_{11}^{i\stackrel{j}{\rightarrow} i'} -J_{12}^{i\stackrel{j}{\rightarrow} i'}(A_{22}^{i})^{-1}A_{21}^i) \\ & + J_{21}^{i\stackrel{j}{\rightarrow} i'} -J_{22}^{i\stackrel{j}{\rightarrow} i'}(A_{22}^{i})^{-1}A_{21}^i ,\\
 R^{i\stackrel{j}{\rightarrow} i'}_{22} =\; & (A_{22}^{i'})^{-1}A_{21}^{i'}J_{12}^{i\stackrel{j}{\rightarrow} i'}+J_{22}^{i\stackrel{j}{\rightarrow} i'}.
\end{split}
\end{equation*}
}

One can then define the reduced order model, formally given by the switched system with single time scale:
\begin{equation}
\label{eq:reducflow}
\dot x(t) = A_0^{\sigma_k} x(t),  \ \forall t\in [t_k,t_{k+1}), \; k\in \N
\end{equation}
with impulsive dynamics:
\begin{equation}\label{eq:reducjump}
x(t_k)=R^{\sigma_{k-1}\stackrel{\nu_{k}}{\rightarrow} \sigma_{k}}_{11} x(t_k^-), \quad \forall k \ge 1.
\end{equation}

The goal of the paper is to investigate the stability of the general singularly perturbed linear hybrid system \eqref{general_fast_slow_dyn}-\eqref{original_reset-map}, or equivalently of 
 \eqref{equation_original}-\eqref{reset-map} or of \eqref{system_fast_slow}-\eqref{reset-map_fast-slow}, for small values of the parameter $\varepsilon$, 
 and its relation to the stability of the reduced order model \eqref{eq:reducflow}-\eqref{eq:reducjump}.
In particular, we aim at characterizing an upper-bound on the minimum dwell-time ensuring stability.

%

\section{Preliminaries} \label{Preliminaries}
In this section, we \blue{provide new} results on the Lyapunov  stability of  singularly perturbed linear systems, which  will be used in the next sections to prove the main results of the paper concerning the stability of \eqref{general_fast_slow_dyn}-\eqref{original_reset-map}.
\blue{The proofs of these results are stated in appendix.}

Let us consider the  singularly perturbed linear system:
\begin{equation}\label{classic_system}
\left\{
\begin{split}
\dot {x}(t)&=A_{11}x(t)+ A_{12}z(t)\\
\varepsilon \dot {z}(t)&=A_{21} x(t) + A_{22}z(t)\\
\end{split}\right.
\end{equation}
where $x(t) \in \R^{n_x}$, $z(t) \in \R^{n_z} $ and $\varepsilon >0$ is a small parameter.
Let us assume that $A_{22}$ is non-singular and proceed with the change of variable
\begin{equation}\label{classic_change}
\left(\begin{array}{c}x(t)\\ y(t)\end{array}\right)=\left(\begin{array}{cc}\I_{n_x} & \0_{n_x,n_z}\\
A_{22}^{-1}A_{21}&  \I_{n_z}\end{array}\right)\left(\begin{array}{c}x(t)\\ z(t)\end{array}\right).
\end{equation}
In the variables $x,y$ the system becomes:
\begin{equation}\label{equation_simple}
\begin{cases}
\dot {x}(t)=A_{0} x(t)+ B_{1} y(t)\\
\varepsilon \dot {y}(t)= A_{22} y(t) + \varepsilon  (B_{2} x(t) + B_{3} y(t))\\
\end{cases}
\end{equation}
where
\begin{align*} A_{0} = A_{11}-A_{12} A_{22}^{-1} A_{21}, \ B_{1}= A_{12}, \\ B_2=A_{22}^{-1} A_{21}A_{0}, \ B_3= A_{22}^{-1} A_{21}A_{12}.\end{align*}
Let us make the following assumption:
\begin{assumption}\label{assum:linear}
$A_0$ and $A_{22}$ are Hurwitz.
\end{assumption}
Under the previous assumption, there exist symmetric positive definite matrices $Q_s \geq \I_{n_x}$, $Q_f \geq \I_{n_z}$ and positive numbers $\lambda_s$ and $\lambda_f$ such that:
\begin{equation*}
\begin{split}
A_0^{\top} Q_s +Q_s A_0^{\top} &\leq -2 \lambda_s Q_s\\
A_{22}^{\top} Q_f +Q_f A_{22}^{\top} &\leq -2 \lambda_f Q_f
\end{split}
\end{equation*}
\blue{Then, let us define $ b_1 =\big\|Q_s^{\frac{1}{2}} B_1Q_f^{-\frac{1}{2}}\big\| ,\ b_2= \big\|Q_f^{\frac{1}{2}}B_2 Q_s^{-\frac{1}{2}}\big\|$ and $ b_3= \big\|Q_f^{\frac{1}{2}}Q_f B_3Q_f^{-\frac{1}{2}}\big\|$.}\\
\blue{The next results are instrumental for our development and their proofs are provided in the Appendix.}

\blue{
\begin{proposition}\label{Proposition1}
Under Assumption~\ref{assum:linear}, $$V(x,y) = x^{\top} Q_s x + y^{\top} Q_f y $$ is a Lyapunov function for system \eqref{equation_simple} for all $ \varepsilon \in (0,\varepsilon_1]$
where 
\begin{equation}\label{epsilon1}
 \varepsilon_1 = \frac{\lambda_f}{\frac{(b_1+b_2)^{2}}{4\lambda_s}+{b_3}}.
\end{equation}
\end{proposition}}

In the following, let us denote \ $W_s(t)=\sqrt{x(t)^{\top}Q_s x(t)}$ and $W_f(t)=\sqrt{y(t)^{\top} Q_f y(t)}$.
\begin{proposition}\label{Proposition2}
Under Assumption~\ref{assum:linear},
let $\varepsilon_1$ be given by \eqref{epsilon1}, then for all $\varepsilon \in (0,\varepsilon_1]$ and $t \geq 0$
$$W_f(t)\leq W_f(0) \e^{-\frac{\lambda_f}{\varepsilon}t}+   \varepsilon  \beta_1 \sqrt{V(0)}  $$
where $\beta_1 = \frac{\sqrt{b_2^2+b_3^2}}{\lambda_f}$.
\end{proposition}

\begin{proposition}
\label{Proposition3}
Under Assumption~\ref{assum:linear}, 
let $\varepsilon_1$ be given by \eqref{epsilon1}, and let \blue{$\varepsilon_2 \in (0, \varepsilon_1] \cap (0,\frac{\lambda_f}{\lambda_s})$}
then for all $\varepsilon \in (0,\varepsilon_2]$ and $t \geq 0$
\begin{equation*}
\begin{split}
W_s(t) \leq \;&  W_s(0) \e^{-\lambda_s t}+  \varepsilon \beta_2 W_f(0) + \varepsilon \beta_3 \sqrt{V(0)} 
\end{split}
\end{equation*}
where $\beta_2=\frac{b_1}{\lambda_f-\varepsilon_2\lambda_s} $ and $\beta_3=\frac{b_1 \beta_1}{\lambda_s}$.
\end{proposition}


%

\section{Stability analysis}\label{Sec:stability}

We now study the stability of system \eqref{system_fast_slow}-\eqref{reset-map_fast-slow} (or equivalently of \eqref{general_fast_slow_dyn}-\eqref{original_reset-map} or of 
 \eqref{equation_original}-\eqref{reset-map}).
In the rest of the paper, we impose the following additional assumption on the singularly perturbed system at hand, related to the stability of the slow and fast dynamics of each mode.
\begin{assumption}\label{assumption_2}
$A_0^{i}$ and $A_{22}^{i}$ are Hurwitz for all $i\in\II$.
\end{assumption}
\blue{
From the previous assumption, we can deduce that there exist symmetric positive definite matrices $Q^{i}_s\geq \I_{n_x}$, $Q_f^{i} \geq \I_{n_z} $, $i \in \II$, and positive numbers $\lambda^i_s$ and $\lambda^i_f$ such that for all $i \in \II$:
\begin{equation*}
\begin{split}
A_0^{i^{\top}} Q^{i}_s +Q^{i}_s A_0^{i} &\leq -2 \lambda^i_s Q^{i}_s\\
A_{22}^{i^{\top}} Q^{i}_f +Q_f^{i} A_{22}^{i} &\leq -2 \lambda^i_f Q_f^{i}
\end{split}
\end{equation*}
We denote  $\lambda_s= \displaystyle \min_{i \in \II}\lambda_s^i$ and $\lambda_f= \displaystyle \min_{i \in \II}\lambda_f^i$.}
For each $i \in \II$, let 
$b_1^i= \big\|(Q^i_s)^{\frac{1}{2}} B_1(Q_f^i)^{-\frac{1}{2}}\big\| $,\ 
$b_2^i= \big\|(Q^i_f)^{\frac{1}{2}}B_2 (Q^i_s)^{-\frac{1}{2}}\big\|$,\ 
$b_3^i=\big\|(Q_f^i)^{\frac{1}{2}}Q_f B_3(Q_f^i)^{-\frac{1}{2}}\big\|$
and $b_j = \displaystyle \max_{i \in \II}b_j^i$ , $j =1,\ldots,3$.

Let $\varepsilon_1$ be given by \eqref{epsilon1}, then it follows from Proposition \ref{Proposition1} that the linear dynamics of  \eqref{system_fast_slow} are all Lyapunov stable, for $\varepsilon \in (0,\varepsilon_1]$. 
Let \blue{$\varepsilon_2 \in (0, \varepsilon_1] \cap (0,\frac{\lambda_f}{\lambda_s})$} and $\beta_1,\ \beta_2,\ \beta_3$ be defined as in Propositions \ref{Proposition2} and \ref{Proposition3}.

The stability analysis of system \eqref{system_fast_slow}-\eqref{reset-map_fast-slow} is carried out using the following functions
\begin{equation*}
\left\{
\begin{array}{l}
W_s(t)=\sqrt{x(t)^{\top}Q^{\sigma_k}_s x(t)} \\
W_f(t)=\sqrt{y(t)^{\top} Q^{\sigma_k}_f y(t)}
\end{array}
\right. ,\; \forall t\in [t_k,t_{k+1}),\; k\in \N.
\end{equation*}
The next result characterizes the variation of $W_s$ and $W_f$ during the continuous dynamics between two events:
\begin{lemma}\label{lemma_1} Under Assumption~\ref{assumption_2},
let \blue{$\varepsilon \in (0,\varepsilon_2]$}, and let $\uptau_k =t_{k+1}-t_k$ for a sequence $(t_k)_{k\geq0} $ of event times. Then for all $k \in \N$,
\begin{align}
\nonumber
W_s(t_{k+1}^-)\leq \;& W_s(t_{k})(\e^{-\lambda_s \uptau_{k}}+\varepsilon\beta_3)  
 +W_f(t_{k}) \varepsilon (\beta_2+\beta_3)\\
 \nonumber
W_f(t_{k+1}^-)\leq \;& W_s(t_{k})\varepsilon\beta_1 + W_f(t_{k})\big(\e^{-\frac{\lambda_f}{\varepsilon}\uptau_{k}}+\varepsilon\beta_1 \big).
\end{align}
\end{lemma}
\begin{IEEEproof}
This is straightforward from Propositions \ref{Proposition2} and \ref{Proposition3} by remarking that $\sqrt{V}\leq W_s + W_f$.
\end{IEEEproof}

In the following we complete the characterization of the variation of $W_s$ and $W_f$ by analyzing their behavior when an event occurs.
Let $\gamma_{11}$, $\gamma_{12}$, $\gamma_{21}$, $\gamma_{22}$ be defined as:
\begin{equation}
\label{eq:gamma}
\begin{split}
\gamma_{11} = \;& \max_{i,i'\in \II,j\in \J}\big\| (Q^{i'}_s)^{\frac{1}{2}} R_{11}^{i\stackrel{j}{\rightarrow}i'} (Q^{i}_s)^{-\frac{1}{2}}\big\|, \\
\gamma_{12} = \;&  \max_{i,i'\in \II,j\in \J}\big\|(Q^{i'}_s)^{\frac{1}{2}} R_{12}^{i\stackrel{j}{\rightarrow}i'} (Q^{i}_f)^{-\frac{1}{2}}\big\|, \\
\gamma_{21} = \;&  \max_{i,i'\in \II,j\in \J}\big\|(Q^{i'}_f)^{\frac{1}{2}} R_{21}^{i\stackrel{j}{\rightarrow}i'} (Q^{i}_s)^{-\frac{1}{2}}\big\|, \\
\gamma_{22} = \;&  \max_{i,i'\in \II,j\in \J}\big\|(Q^{i'}_f)^{\frac{1}{2}} R_{22}^{i\stackrel{j}{\rightarrow}i'} (Q^{i}_f)^{-\frac{1}{2}}\big\|. 
\end{split}
\end{equation}
Then, we have the following result:
\begin{lemma}\label{lemma_2}
Let  a sequence $(t_k)_{k\geq0} $ of event times, then for all $k \ge 1$,
\begin{align*}
W_s(t_{k})\leq \;& \gamma_{11} W_s(t_{k}^-) +\gamma_{12} W_f(t_{k}^-) \\
W_f(t_{k})\leq \;& \gamma_{21}W_s(t_{k}^-)+ \gamma_{22} W_f(t_{k}^-).
\end{align*}
\end{lemma}

\begin{IEEEproof} We prove the first inequality:
\begin{equation*}
\begin{split}
W_s(t_k)=\;& \sqrt{x(t_k)^{\top}Q^{\sigma_k}_s x(t_k)} = \big\| (Q^{\sigma_k}_s)^{\frac{1}{2}}x(t_k) \big\| \\
\le\; & \big\| (Q^{\sigma_k}_s)^{\frac{1}{2}} \big(R_{11}^{\sigma_{k-1}\stackrel{\nu_k}{\rightarrow} \sigma_{k}} x(t_k^-)
+R_{12}^{\sigma_{k-1}\stackrel{\nu_k}{\rightarrow} \sigma_{k}} y(t_k^-)\big)\big\|
 \\
\le \; &  \big\| (Q^{\sigma_k}_s)^{\frac{1}{2}} R_{11}^{\sigma_{k-1}\stackrel{\nu_k}{\rightarrow} \sigma_{k}} x(t_k^-) \big\| \\
& + \big\| (Q^{\sigma_k}_s)^{\frac{1}{2}} R_{12}^{\sigma_{k-1}\stackrel{\nu_k}{\rightarrow} \sigma_{k}} y(t_k^-)\big)\big\| \\
\le \; & \big\| (Q^{\sigma_k}_s)^{\frac{1}{2}} R_{11}^{\sigma_{k-1}\stackrel{\nu_k}{\rightarrow} \sigma_{k}} (Q^{\sigma_{k-1}}_s)^{-\frac{1}{2}} \big\| W_s(t_k^-) \\
&+\big\| (Q^{\sigma_k}_s)^{\frac{1}{2}} R_{12}^{\sigma_{k-1}\stackrel{\nu_k}{\rightarrow} \sigma_{k}} (Q^{\sigma_{k-1}}_f)^{-\frac{1}{2}} \big\| W_f(t_k^-) \\
\le \;& \gamma_{11} W_s(t_{k}^-) +\gamma_{12} W_f(t_{k}^-). 
 \end{split}
\end{equation*}
The second inequality is obtained similarly.
\end{IEEEproof}

In order to keep the notation simple, we introduce the positive matrix parameterized by $\uptau>0$:
\begin{equation*}
M_{\uptau}= \left(\begin{array}{cc}
 \e^{-\lambda_s \uptau}+\varepsilon \beta_3& \varepsilon (\beta_2 +\beta_3)  \\
\varepsilon \beta_1 & \e^{-\frac{\lambda_f}{\varepsilon}\uptau}+ \varepsilon \beta_1
\end{array}\right).
\end{equation*}
Let us also consider the positive matrix
\begin{equation*}
\Gamma = \left(
\begin{array}{cc}
\gamma_{11} & \gamma_{12} \\
\gamma_{21} & \gamma_{22}
\end{array}
\right).
\end{equation*}
\begin{lemma}\label{lemma_3}
Under Assumption~\ref{assumption_2},
let \blue{$\varepsilon \in (0,\varepsilon_2]$}, and let $\uptau_k =t_{k+1}-t_k$ for a sequence $(t_k)_{k\geq0} $ of event times. Then for all $k \in \N$,
\begin{align*}
\left(\begin{array}{c}
W_s(t_{k+1})\\
W_f(t_{k+1})
\end{array}\right)& \leq \Gamma M_{\uptau_k}
\left(\begin{array}{c}
W_s(t_{k})\\
W_f(t_{k})
\end{array}\right).
\end{align*}
\end{lemma}
\begin{IEEEproof}
This is straightforward from Lemmas~\ref{lemma_1} and \ref{lemma_2}.
\end{IEEEproof}

\begin{lemma}
\label{lem:stabschur}
Under Assumption~\ref{assumption_2}, let \blue{$\varepsilon \in (0,\varepsilon_2]$} and let $\uptau^* \ge 0$ such that the positive matrix 
$\Gamma M_{\uptau^*}$ is Schur. Then, for all sequences $(t_k)_{k\geq0} $ of event times satisfying the dwell-time property 
$\uptau_k\ge \uptau^*$, for all $k\in \N$, the system \eqref{system_fast_slow}-\eqref{reset-map_fast-slow} is globally asymptotically stable.
\end{lemma}

\begin{IEEEproof}
From Lemma~\ref{lemma_3}, it follows that for all $k \in \N$,
\begin{equation*}
\left(\begin{array}{c}
W_s(t_{k})\\
W_f(t_{k})
\end{array}\right) \leq \Gamma M_{\uptau_{k-1}} \dots \Gamma M_{\uptau_0}
\left(\begin{array}{c}
W_s(t_{0})\\
W_f(t_{0})\end{array}\right).
\end{equation*}
Remarking that the coefficient of the positive matrix $M_\tau$ are decreasing with respect to $\uptau$, it follows that
\begin{equation*}
\left(\begin{array}{c}
W_s(t_{k})\\
W_f(t_{k})
\end{array}\right) \leq \big(\Gamma M_{\uptau^*}\big)^k
\left(\begin{array}{c}
W_s(t_{0})\\
W_f(t_{0})\end{array}\right).
\end{equation*}
Hence, if the positive matrix $\Gamma M_{\uptau^*}$ is Schur, then both sequences $( W_s(t_{k}))_{k\geq0}$ and $(W_f(t_{k}))_{k\geq0}$ go to $0$, and
the system \eqref{system_fast_slow}-\eqref{reset-map_fast-slow} is globally asymptotically stable.
\end{IEEEproof}

Hence, the stability of system \eqref{system_fast_slow}-\eqref{reset-map_fast-slow} can be investigated by studying the spectral properties of the positive matrix 
$\Gamma M_{\uptau^*}$. \blue{Let us remark that values $\uptau^*$ such that $\Gamma M_{\uptau^*}$ is Schur provide  upper bounds on the minimal dwell-time between two events that ensures the stability of the singularly perturbed linear hybrid system. In the following, we establish sufficient conditions for deriving such values $\uptau^*$. The proofs are provided in appendix.}
We consider three distinct cases depending on the value of parameter $\gamma_{11}$ defined in~\eqref{eq:gamma}. 

\subsection{Case 1:  \blue{$\gamma_{11}>1$}}

\begin{theorem} \label{theorem_final}
Under Assumption~\ref{assumption_2}, let $\gamma_{11}>1$. Then, there exists $\varepsilon_1^*>0$ and a function $\eta_1: (0,\varepsilon_1^*)\rightarrow \R^+$ with $\eta_1(\varepsilon)=\mathcal O (\varepsilon)$,
such that for all $\varepsilon \in (0,\varepsilon_1^*)$,
for all sequences $(t_k)_{k\geq0} $ of event times satisfying a dwell-time property 
$\uptau_k\ge \uptau^*$, for all $k\in \N$, with
\begin{equation*}
\uptau^* > \frac{\ln(\gamma_{11})}{\lambda_s}+\eta_1(\varepsilon),
\end{equation*}
the system \eqref{system_fast_slow}-\eqref{reset-map_fast-slow} is globally asymptotically stable.
\end{theorem}

Theorem~\ref{theorem_final} shows that a dwell-time ensuring stability of the singularly perturbed switched impulsive system \eqref{system_fast_slow}-\eqref{reset-map_fast-slow} can be written as the sum of a constant part 
$\frac{\ln(\gamma_{11})}{\lambda_s}$ and of a function $\eta_1(\varepsilon)$, which goes to $0$ as fast as $\varepsilon$ when $\varepsilon$ goes to $0$.
\blue{Interestingly, the constant part only depends on $\lambda_s$ and $\gamma_{11}$, which can be determined only from the reduced order model \eqref{eq:reducflow}-\eqref{eq:reducjump}.
Moreover, we will show in Section~\ref{sec:red} that \eqref{eq:reducflow}-\eqref{eq:reducjump} is globally asymptotically stable for all switching signals with dwell-time  $\tau^*>\frac{\ln(\gamma_{11})}{\lambda_s}$.}


\subsection{Case 2: $\gamma_{11}=1$}

When $\gamma_{11}= 1$, two cases can be distinguished depending on whether $\gamma_{12}\ne 0$ or $\gamma_{12}=0$. In the latter case, this means that the slow variable $x$ is not influenced by the fast variable $y$ through jumps \blue{(i.e. $J_{12}^{i\stackrel{j}{\rightarrow} i'}=0$ for all $i,i' \in \II$, $j\in \J$)}.

\begin{theorem}\label{theorem_final2}
Under Assumption~\ref{assumption_2}, let $\gamma_{11}= 1$ \blue{and $\gamma_{12}\neq0$.} Then, there exists $\varepsilon_2^*>0$ and a function $\eta_2: (0,\varepsilon_2^*)\rightarrow \R^+$ with $\eta_2(\varepsilon)=\mathcal O (\varepsilon)$,
such that for all $\varepsilon \in (0,\varepsilon_2^*)$,
for all sequences $(t_k)_{k\geq0} $ of event times satisfying a dwell-time property 
$\uptau_k\ge \uptau^*$, for all $k\in \N$, with
\begin{equation*}
\uptau^* > \frac{-\varepsilon}{\lambda_f}\ln(\varepsilon)+\eta_2(\varepsilon),
\end{equation*}
the system \eqref{system_fast_slow}-\eqref{reset-map_fast-slow} is globally asymptotically stable.
\end{theorem}

\begin{theorem}\label{theorem_final3}
Under Assumption~\ref{assumption_2}, let $\gamma_{11}= 1$ and $\gamma_{12}=0$. Then, there exists $\varepsilon_3^*>0$ and a function $\eta_3: (0,\varepsilon_3^*)\rightarrow \R^+$ with $\eta_3(\varepsilon)=\mathcal O (\varepsilon)$,
such that for all $\varepsilon \in (0,\varepsilon_3^*)$,
for all sequences $(t_k)_{k\geq0} $ of event times satisfying a dwell-time property 
$\uptau_k\ge \uptau^*$, for all $k\in \N$, with
\begin{equation*}
\uptau^* > \eta_3(\varepsilon),
\end{equation*}
the system \eqref{system_fast_slow}-\eqref{reset-map_fast-slow} is globally asymptotically stable.
\end{theorem}


The previous theorems show that when $\gamma_{11} = 1$, the minimum dwell-time ensuring stability of the singularly perturbed switched impulsive system \eqref{system_fast_slow}-\eqref{reset-map_fast-slow} goes to $0$ as fast as $-\varepsilon \ln(\varepsilon)$ or $\varepsilon$
when $\varepsilon$ goes to $0$. It is interesting to remark that in that case, as we will show in Section~\ref{sec:red}, 
the reduced order system \eqref{eq:reducflow}-\eqref{eq:reducjump} is globally asymptotically stable for all switching signals without any dwell-time condition. It is also noticeable that when $\gamma_{12}\ne 0$, the dwell-time is larger (by a factor of order $-\ln(\varepsilon)$) than when $\gamma_{12}=0$. In the former case, more time is needed to stabilize the fast variable $y$ so that it does not destabilize the slow variable through the impulsive dynamics.


\subsection{Case 3: $\gamma_{11} <1$}

When $\gamma_{11}<1$ one can again distinguish two cases depending on the values of other parameters $\gamma_{12}$, $\gamma_{21}$ and $\gamma_{22}$.

\begin{theorem}\label{theorem_final4}
Under Assumption~\ref{assumption_2}, let $\gamma_{11}< 1$. Then, there exists $\varepsilon_4^*>0$ and a function $\eta_4: (0,\varepsilon_4^*)\rightarrow \R^+$ with $\eta_4(\varepsilon)=\mathcal O (\varepsilon)$,
such that for all $\varepsilon \in (0,\varepsilon_4^*)$,
for all sequences $(t_k)_{k\geq0} $ of event times satisfying a dwell-time property 
$\uptau_k\ge \uptau^*$, for all $k\in \N$, with
\begin{equation*}
\uptau^* > \eta_4(\varepsilon),
\end{equation*}
the system \eqref{system_fast_slow}-\eqref{reset-map_fast-slow} is globally asymptotically stable.
\end{theorem}

\begin{theorem}\label{theorem_final5}
Under Assumption~\ref{assumption_2}, let $\gamma_{11}< 1$, $\gamma_{22}<1$ and $\frac{\gamma_{12}\gamma_{21}}{(1-\gamma_{11})(1-\gamma_{22})}<1$. Then, there exists $\varepsilon_5^*>0$,
such that for all $\varepsilon \in (0,\varepsilon_5^*)$,
for all sequences $(t_k)_{k\geq0} $ of event times, 
the system  \eqref{system_fast_slow}-\eqref{reset-map_fast-slow}  is globally asymptotically stable.
\end{theorem}


The previous theorems show that when $\gamma_{11} < 1$, the minimum dwell-time ensuring stability of the singularly perturbed switched impulsive system \eqref{system_fast_slow}-\eqref{reset-map_fast-slow} is either equal to $0$ or goes to $0$ as f	ast as $\varepsilon$ when $\varepsilon$ goes to $0$. We will show in the next section that in that case,  
the reduced order system \eqref{eq:reducflow}-\eqref{eq:reducjump} is globally asymptotically stable for all switching signals without any dwell-time condition.


\begin{table*}[!t]
\caption{\label{tab} Summary of the main results of the paper establishing dwell-time conditions for the stability of the singularly perturbed hybrid system \eqref{system_fast_slow}-\eqref{reset-map_fast-slow} and of the the reduced order system \eqref{eq:reducflow}-\eqref{eq:reducjump}.}
\begin{center}
\begin{tabular}{|c|c|c|c|}
\hline
$\gamma_{11}$ & $\gamma_{12},\; \gamma_{21},\; \gamma_{22}$ & dwell-time condition for \eqref{system_fast_slow}-\eqref{reset-map_fast-slow} & dwell-time condition for \eqref{eq:reducflow}-\eqref{eq:reducjump} \\
\hline
$\gamma_{11}>1$ & -- & $\uptau^* > \frac{\ln(\gamma_{11})}{\lambda_s} + \mathcal O(\varepsilon)$& $\uptau^* > \frac{\ln(\gamma_{11})}{\lambda_s}$ \\
\hline
\multirow{2}*{$\gamma_{11}=1$} & -- & $\uptau^* > -\frac{\varepsilon}{\lambda_f}\ln(\varepsilon) + \mathcal O(\varepsilon)$& \multirow{4}*{$\uptau^* \ge 0$} \\
\cline{2-3}
& $\gamma_{12}=0$ & \multirow{2}*{$\uptau^* > \mathcal O(\varepsilon)$} &\\
\cline{1-2}
\multirow{2}*{$\gamma_{11}<1$} & -- & &  \\
\cline{2-3}
& $\gamma_{22}<1$, $\frac{\gamma_{12}\gamma_{21}}{(1-\gamma_{11})(1-\gamma_{22})}<1$
 & $\uptau^* \ge 0$ &\\
\hline
\end{tabular}
\end{center}
\end{table*}

\subsection{Stability of reduced order system}
\label{sec:red}

\blue{
It is interesting to remark that in the previous results, the upper bound on the minimum dwell-time ensuring stability of system  \eqref{system_fast_slow}-\eqref{reset-map_fast-slow}  can be seen as the sum of two terms. The first term is independent of the parameter $\varepsilon$, its value is $0$ when $\gamma_{11}\le 1$ and $\frac{\ln{\gamma_{11}}}{\lambda_s}$ when $\gamma_{11}>1$. 
The second term depends on the parameter $\varepsilon$ and goes to $0$ when $\varepsilon$ goes to $0$. In this section, we show that an interpretation of the first term can be given in terms of the reduced-order system  \eqref{eq:reducflow}-\eqref{eq:reducjump}, since it provides an upper bound on the minimum dwell-time guaranteeing stability for that system.}

\begin{proposition}\label{pro:red1}  
Under Assumption~\ref{assumption_2}, let $\gamma_{11}>1$. Then, for all sequences $(t_k)_{k\geq0} $ of event times satisfying a dwell-time property 
$\uptau_k\ge \uptau^*$ with $\uptau^* > \frac{\ln(\gamma_{11})}{\lambda_s}$, for all $k\in \N$, the reduced order system \eqref{eq:reducflow}-\eqref{eq:reducjump} is globally asymptotically stable.
\end{proposition}

\begin{IEEEproof} We  consider the function $W_s$ given by $W_s(t)=\sqrt{x(t)^\top Q_s^{\sigma_k} x(t)}$, for all $t\in [t_k,t_{k+1})$, $k\in \N$. 
By Assumption~\ref{assumption_2}, it follows that for all $k\in \N$, $W_s(t_{k+1}^-)\le W_s(t_k) e^{-\lambda_s \uptau_k} $. Moreover, from the definition of $\gamma_{11}$ in \eqref{eq:gamma}, it follows that  for all $k\in \N$, $W_s(t_{k+1})\le W_s(t_{k+1}^-) \gamma_{11}$. Hence,
\begin{equation}
\label{eq:lyapred}
\forall k\in \N,\; W_s(t_{k+1})\le W_s(t_k) \gamma_{11} e^{-\lambda_s \uptau_k}. 
\end{equation}
Then, since for all $k\in \N$, $\uptau_k\ge \uptau^*$ with $\uptau^* > \frac{\ln(\gamma_{11})}{\lambda_s}$, it follows that $W_s(t_k)$ goes to $0$ as $k$ goes to $+\infty$ and
the reduced order system \eqref{eq:reducflow}-\eqref{eq:reducjump} is globally asymptotically stable.
\end{IEEEproof}

\begin{proposition}\label{pro:red2}  
Under Assumption~\ref{assumption_2}, let $\gamma_{11}\le 1$. Then, for all \blue{unbounded}  sequences $(t_k)_{k\geq0} $ of event times, the reduced order system \eqref{eq:reducflow}-\eqref{eq:reducjump} is globally asymptotically stable.
\end{proposition}

\begin{IEEEproof} From \eqref{eq:lyapred}, it follows from  \blue{$\gamma_{11}\le 1$} that for all $k\in \N$, $W_s(t_{k+1})\le W_s(t_k) e^{-\lambda_s \uptau_k}$. 
Therefore, for all $k\in \N$, $W_s(t_k) \le W_s(0) e^{-\lambda_s t_k}$. \blue{Since $(t_k)_{k\geq0} $ is unbounded}, $t_k$ goes to $+\infty$ and therefore $W_s(t_k)$ goes to $0$. 
Thus, the reduced order system \eqref{eq:reducflow}-\eqref{eq:reducjump} is globally asymptotically stable.
\end{IEEEproof}

The previous propositions show that the dwell-time condition established for the singularly perturbed hybrid system \eqref{system_fast_slow}-\eqref{reset-map_fast-slow}
in Theorems~\ref{theorem_final}, \ref{theorem_final2}, \ref{theorem_final3}, \ref{theorem_final4} and \ref{theorem_final5} coincides when $\varepsilon$ goes to $0$ with the dwell-time condition of the reduced order system given in \eqref{eq:reducflow}-\eqref{eq:reducjump}. 
\blue{Since $\varepsilon$ is assumed to be small, it appears that the main source of conservatism in the dwell-time estimates for the singularly perturbed hybrid system \eqref{system_fast_slow}-\eqref{reset-map_fast-slow} comes from the dwell-time estimates of the reduced order system.}
Table~\ref{tab} summarizes the main results of the paper.

%

\section{Extension to the case of time varying state's dimensions vectors}\label{extension}

In this section, we briefly explain how we can use the previous results for the analysis of system \eqref{general_fast_slow_dyn}-\eqref{original_reset-map} in the case of time-varying dimensions of the slow and fast state vectors. We recall that for all $t\in [t_k,t_{k+1})$ the state vector $\x(t)$ of system  \eqref{general_fast_slow_dyn} is of dimension $n_i$ when $\sigma_k=i$. Let us also recall that matrices $\DD^i,\ i\in\II$ were  introduced in section \ref{Problem formulation} to define the dynamics \eqref{general_fast_slow_dyn}. For all $i\in\II$ we consider that $n_{zi}\in\N$ is the number of  elements of $\DD^i$ that are equal with $\varepsilon$ and $n_{xi}=n_i-n_{zi}$. In other words, when $\sigma_k=i$, $n_i, n_{zi}$ and $n_{xi}$ are the dimensions of the state vector, fast variables vector and slow variables vector, respectively. Furthermore, let us introduce $n_z=\displaystyle\max_{i\in\II}n_{zi}$, $n_x=\displaystyle\max_{i\in\II}n_{xi}$ and $n=n_x+n_z$.

With the notation introduced above we define the following augmented system:

\begin{equation}\label{augmented_fast_slow_dyn}
\left\{\begin{split}
 \DD^{\sigma_k}\dot{\x}(t)=\AA^{\sigma_k}\x(t), \\
\bar{\DD}^{\sigma_k}\dot{\bar{\x}}(t)=-\lambda \bar{\x}(t) , 
\end{split}\right. \quad \forall t\in [t_k,t_{k+1}), \; k\in \N
\end{equation}
where $\bar{\DD}^{i}\in\R^{(n-n_{i})\times (n-n_{i})}$ is defined similarly to $\DD^{i}$ as a diagonal matrix with $\varepsilon$ or $1$ diagonal elements, used to select the fast and slow variable from the components of the artificial state vector $\bar{\x}$. To be precise, for all $i\in\II$ we consider $\bar{\DD}^{i}$ having $n_z-n_{zi}$ diagonal elements equal to $\varepsilon$. Consequently, the augmented vector $\left(\begin{smallmatrix} \x(t)\\
\bar{\x}(t)\end{smallmatrix}\right)$ has an invariant number of slow and fast components which is $n_x$ and $n_z$, respectively. Therefore, \eqref{augmented_fast_slow_dyn} is of the form \eqref{general_fast_slow_dyn} and the dimension of its state vector as well as the number of its slow and fast variables are constant. The parameter $\lambda$ is a positive number that can be chosen greater than $\lambda_s$ and $\lambda_f$ in order to make the continuous dynamics of the auxiliary variable $\bar \x(t)$ converge faster than that of the variable $\x(t)$.

%

Secondly, we define a jump map for the augmented vector as follows:
\begin{equation}\label{extended_jump-map}
\left\{\begin{array}{lll}
\x(t_k) &= &\JJ^{\nu_k} \x(t_k^-) \\
\bar{\x}(t_k)& =& 0
\end{array}\right.
\end{equation}
The auxiliary variable is set to $0$ at jumps so that the discrete dynamics of the auxiliary variable $\bar \x(t)$ converge faster than that of the variable $\x(t)$.
It is clear that the augmented system \eqref{augmented_fast_slow_dyn}-\eqref{extended_jump-map} is globally asymptotically stable if and only if the orignal system  \eqref{general_fast_slow_dyn}-\eqref{original_reset-map} is.
Then, the stability analysis of \eqref{augmented_fast_slow_dyn}-\eqref{extended_jump-map} can be carried out as shown on the previous sections.

\section{Illustration on stability analysis of scalar fast and slow dynamics}\label{sec:scalar}
\subsection{Dwell-time analysis}
This section aims to illustrate the previous analysis on a low dimensional system. We consider a linear singularly perturbed switched system with scalar slow and fast variables. Moreover, we consider that $\II=\{1,2\}$. The objective is to analyze the stability of the system under the assumption that after each switch the slow variable becomes fast and vice-versa. To be more precise let $0=t_0 < t_1 < \dots$ be the sequence of discrete instants where a switch takes place and consider the following dynamics:
\begin{equation}\label{mode1}
\left\{\begin{array}{l}\dot{u}(t)=a_1u(t)+b_1v(t)\\
\varepsilon\dot{v}(t)=c_1u(t)+d_1v(t)\end{array}\right. \ t\in[t_{2k},t_{2k+1}), k\in\N
\end{equation}
and
\begin{equation}\label{mode2}
\left\{\begin{array}{l}
\varepsilon\dot{u}(t)=a_2u(t)+b_2v(t)\\
\dot{v}(t)=c_2u(t)+d_2v(t)\end{array}\right. \ t\in[t_{2k+1},t_{2k+2}), k\in\N
\end{equation}
The dynamics \eqref{mode1}-\eqref{mode2} above can be written in the compact form \eqref{general_fast_slow_dyn} by using $\x=(u,v)^\top$ and the matrices \[\DD^1=\left(\begin{array}{cc}1 & 0\\ 0 & \varepsilon\end{array}\right),\quad \DD^2=\left(\begin{array}{cc}\varepsilon & 0\\ 0 & 1\end{array}\right)\] and \[\AA^1=\left(\begin{array}{cc}a_1 & b_1\\ c_1 & d_1\end{array}\right),\quad \AA^2=\left(\begin{array}{cc}a_2 & b_2\\ c_2 & d_2\end{array}\right).\] 
Introducing the permutation matrices $S_1=\I_2$ and $S_2=\left(\begin{smallmatrix}0 & 1\\ 1 & 0\end{smallmatrix}\right)$ we can define the change of variable \eqref{var-transf} as 
\[\left\{\begin{split}
(x,z)^\top&=S_1\x,  \ t\in[t_{2k},t_{2k+1}), k\in\N, \\ 
(x,z)^\top&=S_2\x,  \ t\in[t_{2k+1},t_{2k+2}), k\in\N.
\end{split}\right.\]
It is worth notting that no jump occurs in the $\x$ variable meaning that $\J=\{1\}$ and $\JJ^1=\I_2$ in \eqref{original_reset-map}. However, it can be seen that the dynamics \eqref{mode1}-\eqref{mode2} expressed in $(x,z)^\top$ variable is an impulsive one.  Precisely, $\J=\{1\}$ but following \eqref{dyn-transf} one obtains that 
\blue{$J^{1\stackrel{1}{\rightarrow} 2}=J^{2\stackrel{1}{\rightarrow} 1}=S_2$}.

Summarizing we can rewrite system \eqref{mode1}-\eqref{mode2} in the form 
\[
\left(\begin{array}{c}
\dot {x}(t)\\
\varepsilon \dot {z}(t)\\
\end{array}\right)
=A^{\sigma_k}
\left(\begin{array}{c}
x(t)\\
z(t)
\end{array}\right), \ \forall t\in [t_k,t_{k+1}), \; k\in \N
\]
with impulsive dynamics:
\blue{
\[
\left(\begin{array}{c}
x(t_k)\\
z(t_k)
\end{array}\right)=J^{\sigma_{k-1}\stackrel{\nu_{k}}{\rightarrow} \sigma_{k}}  \left(\begin{array}{c}
x(t_k^-)\\
z(t_k^-)
\end{array}\right), \quad \forall k \ge 1
\]
}
where $\sigma_k\in \II=\{1,2\}$, $\nu_k \in \J=\{1\}$, \blue{$J^{1\stackrel{1}{\rightarrow} 2}=J^{2\stackrel{1}{\rightarrow} 1}=S_2$}, $A^1=S_1\AA^1S_1^\top=\AA^1$ and $A^2=S_2\AA^2S_2^\top=\left(\begin{smallmatrix}d_2 & c_2\\ b_2 & a_2\end{smallmatrix}\right)$. 

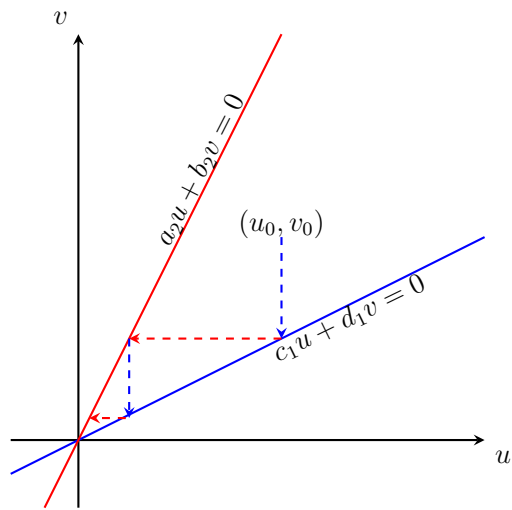
\begin{figure}[h]
\begin{center}
\begin{tikzpicture}[scale = 0.9, transform shape, thick]
\tikzstyle{fleche}=[->, >=stealth,thick,rounded corners=4pt]

\draw[fleche] (-1,0) -- (6,0) node[anchor=north west] {$u$};
\draw[fleche] (0,-1) -- (0,6) node[anchor=south east] {$v$};
\draw[blue] (-1,-0.5) --(6,3);
\draw[red] (-0.5,-1) --(3,6);
\node[rotate=63] at (1.8,4) {$a_2u+b_2v=0$};
\node[rotate=26] at (4,1.8) {$c_1u+d_1v=0$};
\draw[fleche,blue,dashed] (3,3) -- (3,1.5);
\draw[fleche,red,dashed] (3,1.5) -- (0.75,1.5);
\draw[fleche,blue,dashed] (0.75,1.5) -- (0.75,0.326);
\draw[fleche,red,dashed] (0.7,0.325) -- (0.1625,0.325);
\node at (3,3.2) {$(u_0,v_0)$};
\end{tikzpicture}
\end{center}
\caption{In blue the slow manifold associated with \eqref{mode1} and in red the  slow manifold associated with \eqref{mode2} when $\frac{|c_1b_2|}{|d_1a_2|}<1$. The dashed lines represent the asymptotic behavior of the overall system with initial state $(u_0,v_0)$ when $\varepsilon\rightarrow0$ and no dwell-time (or $\mathcal O (\varepsilon)$ dwell-time) is imposed. It can be seen that system \eqref{mode1}-\eqref{mode2} is asymptotically stable for any switching rule.}\label{fig_num1}
\end{figure}

The time dependent change of coordinates \eqref{y_change} is expressed as:
\begin{equation}\label{y_change_scalar}
\begin{split}
&y(t)=z(t)+\frac{c_1}{d_1}x(t),\ t\in [t_k,t_{k+1}), \; k\in \N,\\
&y(t)=z(t)+\frac{b_2}{a_2}x(t),\  t\in[t_{2k+1},t_{2k+2}), k\in\N.
\end{split} 
\end{equation}
\blue{
Assumption \ref{assumption_2} simply requires that 
\[\left\{\begin{split}A_0^1=a_1-\frac{b_1c_1}{d_1}<0, \quad A_{22}^1=d_1<0, \\
A_0^2=d_2-\frac{b_2c_2}{a_2}<0,\quad A_{22}^2=a_2<0.\end{split}\right.\]
Then,  $Q_s^i,\ Q_f^i$, $i\in \II$ can be chosen as any positive scalars and it is easy to check that
$$
\lambda_s=\min\left(\frac{b_1c_1}{d_1}-a_1,\frac{b_2c_2}{a_2}-d_2
 \right),\; \lambda_f=\min(-d_1,-a_2).
$$
In our analysis, an important role is played by the values $R_{11}^{1\stackrel{1}{\rightarrow} 2}$ and $R_{11}^{2\stackrel{1}{\rightarrow} 1}$, which determine the value of $\gamma_{11}$, which in turn (see Theorems \ref{theorem_final}-\ref{theorem_final5} and Table \ref{tab}) allows concluding wether the required dwell time approaches $0$ when $\varepsilon$ goes to $0$.
Therefore it is worth to explicit that:
\[
R_{11}^{1\stackrel{1}{\rightarrow} 2}=-\frac{c_1}{d_1},\quad R_{11}^{2\stackrel{1}{\rightarrow} 1}=-\frac{b_2}{a_2}.
\]
Furthermore, following \eqref{eq:gamma} one has 
$$
\gamma_{11}=\max\left(\left| \frac{qc_1}{d_1}\right| ,  \left|\frac{b_2}{qa_2}\right| \right)
$$
where $q=\sqrt{\frac{Q_s^2}{Q_s^1}}$. For our analysis, it is desirable to have $\gamma_{11}$ as small as possible, it is minimal when $q=\sqrt{ \frac{|d_1b_2|}{|c_1a_2|}}$ and in that case
$$
\gamma_{11}=\sqrt{ \frac{|c_1b_2|}{|d_1a_2|}}.
$$
Then, $\gamma_{11}<1$ if and only if $\frac{|c_1b_2|}{|d_1a_2|}<1$
and following Theorem \ref{theorem_final4},  a dwell-time of order $O(\varepsilon)$ is sufficient to stabilize the system. When $\varepsilon\rightarrow0$ it yields that the switching system given by the two slow manifolds of \eqref{mode1}-\eqref{mode2} is stable whatever is the considered switching rule (\ie no dwell-time required). This result is illustrated in Fig. \ref{fig_num1} which takes into account that the two slow manifolds of \eqref{mode1}-\eqref{mode2} are the lines:
\[
c_1u(t)+d_1v(t)=0
\mbox{ and } 
a_2u(t)+b_2v(t)=0.
\] 
It is noteworthy that $\frac{|c_1b_2|}{|d_1a_2|}<1$ essentially says that the slope of the slow manifold associated with \eqref{mode1} is smaller than the slope of  the slow manifold associated with \eqref{mode2}.}

\begin{figure}[h]
\begin{center}
\begin{tikzpicture}[scale = 0.9, transform shape, thick]

\tikzstyle{fleche}=[->, >=stealth,thick,rounded corners=4pt] 
\draw[fleche] (-1,0) -- (6,0) node[anchor=north west] {$u$};
\draw[fleche] (0,-1) -- (0,6) node[anchor=south east] {$v$};
\draw[red] (-1,-0.5) --(6,3);
\draw[blue] (-0.5,-1) --(3,6);
\node[rotate=63] at (1.8,4) {$c_1u+d_1v=0$};
\node[rotate=26] at (4,1.8) {$a_2u+b_2v=0$};
\draw[fleche,blue,dashed] (1,1) -- (1,2);
\draw[fleche,red,dashed] (1,2) -- (4,2);
\draw[fleche,blue,dashed] (4,2) -- (4,6);
\node at (1,0.9) {$(u_0,v_0)$};
\end{tikzpicture}
\end{center}
\caption{In blue the slow manifold associated with \eqref{mode1} and in red the  slow manifold associated with \eqref{mode2} when  $\frac{|c_1b_2|}{|d_1a_2|}>1$. The dashed lines represent the asymptotic behavior of the overall system with initial state $(u_0,v_0)$ when $\varepsilon\rightarrow0$ and no dwell-time (or $\mathcal O (\varepsilon)$ dwell-time) is imposed. It is illustrated that in this case  a dwell-time of order $\frac{\ln(\gamma_{11})}{\lambda_s}+\mathcal O (\varepsilon)$ has to be imposed in order to guarantee the system's stability.}\label{fig_num2}
\end{figure}
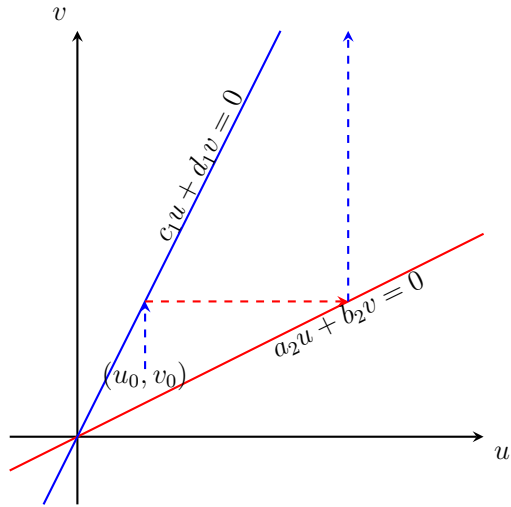

\blue{Reversely,  $\gamma_{11}>1$ if and only if $\frac{|c_1b_2|}{|d_1a_2|}>1$, meaning that the slope of the slow manifold associated with \eqref{mode2} is smaller than the slope of  the slow manifold associated with \eqref{mode1}. In this case we use Theorem \ref{theorem_final}  to deduce that a dwell-time of order $\frac{\ln(\gamma_{11})}{\lambda_s}+\mathcal O (\varepsilon)$
 is required. In absence of dwell-time we can see  in Fig. \ref{fig_num2}  that the switching system given by the two slow manifolds of \eqref{mode1}-\eqref{mode2} is unstable. }

\subsection{Numerical examples}\label{sec: example}

In this section we provide a numerical illustration of the previous results. Let us reconsider system \eqref{mode1}-\eqref{mode2} when the state matrices take the following numerical values:  \begin{equation}\label{num-eval}\AA^1=\left(\begin{array}{cc}-1 & 0.5\\ -1 & -2\end{array}\right),\quad \AA^2=\left(\begin{array}{cc}-2.5 & -2\\ 3 & 1\end{array}\right).\end{equation} 


\blue{
Assumption \ref{assumption_2} holds since  \[\left\{\begin{split}a_1-\frac{b_1c_1}{d_1}=-1.25<0  \quad d_1=-2<0, \\
d_2-\frac{b_2c_2}{a_2}=-1.4<0 \quad a_2=-2.5<0.\end{split}\right.\]
Then, $\lambda_s=1.25$ and $\lambda_f=2$ for any choice of positive scalars  $Q_s^i,\ Q_f^i$, $i\in \II$.
Letting $q=\sqrt{\frac{Q_s^2}{Q_s^1}} =\sqrt{ \frac{|d_1b_2|}{|c_1a_2|}}=2\sqrt{ \frac{2}{5}}$ we obtain $
\gamma_{11}=\sqrt{ \frac{|c_1b_2|}{|d_1a_2|}}=\sqrt{ \frac{2}{5}}<1$.
Therefore, according to Theorem  \ref{theorem_final4}, the minimum stabilizing dwell time is in $\mathcal O(\varepsilon)$.}

Let $\varepsilon = 10^{-3} $ and the initial condition $\x_0=(2,\ 1)$. Using Theorem  \ref{theorem_final4}  one deduces that the required dwell-time for the stability of system \eqref{mode1}-\eqref{mode2} is $6.16 \cdot 10^{-4} = O(\varepsilon)$. 

 \begin{figure}[!h]
\begin{center}
 \includegraphics[scale=0.33]{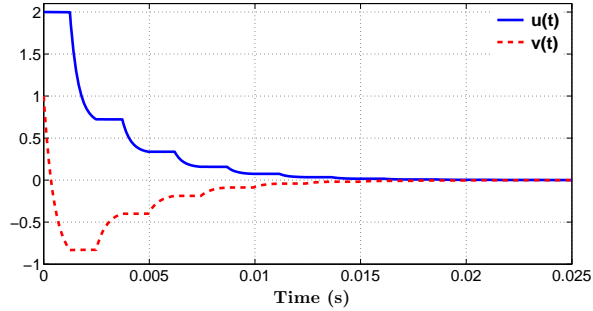}
  \caption{State's trajectory for  \eqref{mode1}-\eqref{mode2} with $\AA_1, \AA_2$ defined by \eqref{num-eval} and $6.16 \cdot 10^{-4} = O(\varepsilon)$}
  \end{center}
 \end{figure}
 
\begin{figure}[!h]
\centering \includegraphics[scale=0.35]{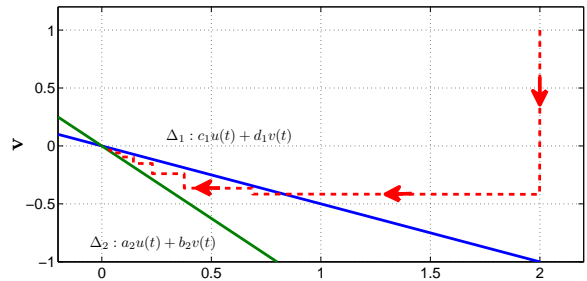}\vspace{-0.2cm}
\caption{State's trajectory in $(u,v)$- plane for \eqref{mode1}-\eqref{mode2} with $\AA_1, \AA_2$ defined by \eqref{num-eval} and  $t_{k+1}-t_k=\tau = 6.16 \cdot 10^{-4} = O(\varepsilon) \ sec$}\label{manifold3}
 \end{figure}

The two slow manifolds of the system are respectively:
\[
\left\{\begin{split}
&-u(t)-2v(t)=0\\
&-2.5 u(t)-2 v(t)=0.\end{split}\right.
\]
The behavior of the system's trajectory in the $(u,v)$- plane is plot in Fig. \ref{manifold3}.
\vspace{-0.2cm}

\begin{figure}[!h]
\centering \includegraphics[scale=0.35]{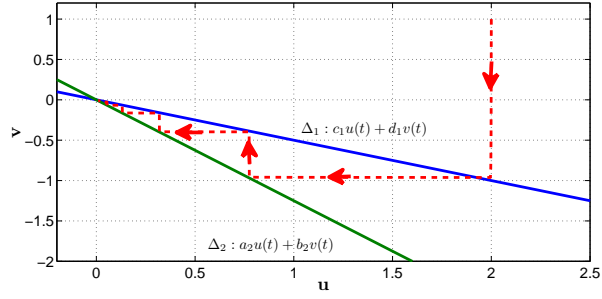}\vspace{-0.2cm}
\caption{State's trajectory in $(u,v)$- plane for \eqref{mode1}-\eqref{mode2} with $\AA_1, \AA_2$ defined by \eqref{num-eval} and  $t_{k+1}-t_k=\tau = 2 \cdot 10^{-3} \ sec$}\label{manifold4}
 \end{figure}

\begin{figure}[!h]
\begin{center}
 \includegraphics[scale=0.35]{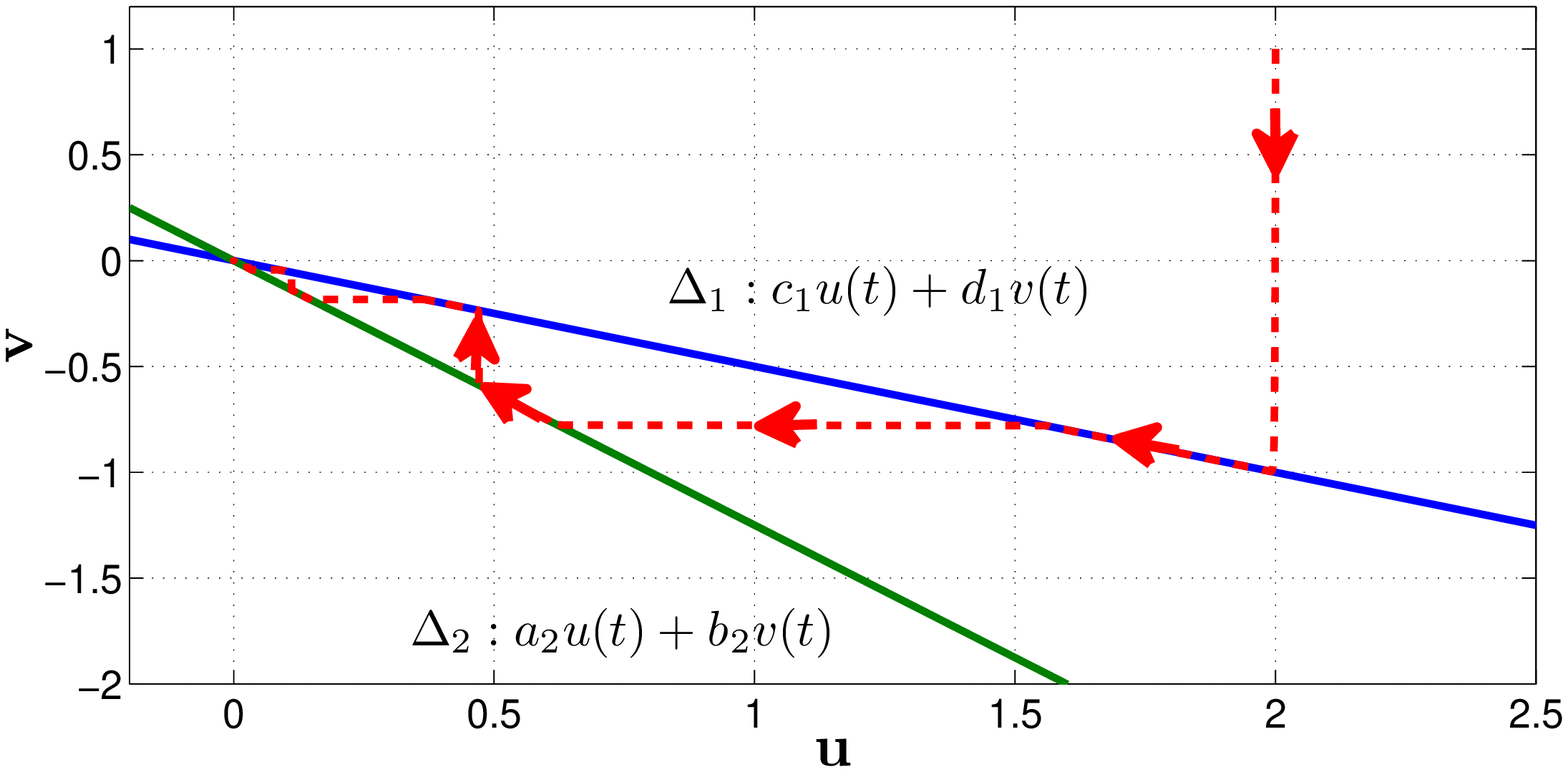}
  \caption{State's trajectory in $(u,v)$- plane for \eqref{mode1}-\eqref{mode2} with $\AA_1, \AA_2$ defined by \eqref{num-eval} and  $t_{k+1}-t_k=\tau = 0.2 \ sec$}\label{Traj2}
  \end{center}
 \end{figure}

It is interesting to note in Fig. \ref{manifold3} that events occurs very fast and far from the origine the system's trajectory approaches the slow manifolds without reaching them. As illustrated in Fig. \ref{manifold4} and Fig. \ref{Traj2} below the system's behavior is not deteriorated when a larger dwell-time condition $t_{k+1}-t_k=\tau=2\cdot10^{-3}\ sec$ or $t_{k+1}-t_k=\tau=0.2\ sec$ is imposed. Moreover, increasing the dwell-time allows the system to reach the slow manifolds and eventually slide on them.
\vspace{-0.2cm}
 
Let us now consider another choice for the state matrices $\AA_1, \AA_2$  in \eqref{mode1}-\eqref{mode2}.
In the following we define:
\begin{equation}\label{num-eval2}
\AA^1=\left(\begin{array}{cc}-1 & 0.5\\ -3 & -2\end{array}\right),\quad \AA^2=\left(\begin{array}{cc}-2.5 & -4\\ 1 & 0.5\end{array}\right).\end{equation} 

Again, one can easily observe that Assumption \ref{assumption_2} holds: \[\left\{\begin{split}a_1-\frac{b_1c_1}{d_1}=-1.75<0,  \quad d_1=-2<0, \\
d_2-\frac{b_2c_2}{a_2}=-1.1<0, \quad a_2=-2.5<0.\end{split}\right.\]
\blue{
Then, $\lambda_s=1.1$ and $\lambda_f=2$ for any choice of positive scalars  $Q_s^i,\ Q_f^i$, $i\in \II$.
Letting $q=\sqrt{\frac{Q_s^2}{Q_s^1}} =\sqrt{ \frac{|d_1b_2|}{|c_1a_2|}}={ \frac{4}{\sqrt {15}}}$ we obtain $
\gamma_{11}=\sqrt{ \frac{|c_1b_2|}{|d_1a_2|}}=2\sqrt{ \frac{3}{5}}>1$.
Therefore, according to Theorem  \ref{theorem_final}, an upper bound on the minimum stabilizing dwell time is given by $\frac{ln(\gamma_{11})}{\lambda_s}+\mathcal O(\varepsilon)$ where,
in the present case, $\frac{ln(\gamma_{11})}{\lambda_s}=0.40sec$.
}

\begin{figure}[!h]
\centering \includegraphics[scale=0.33]{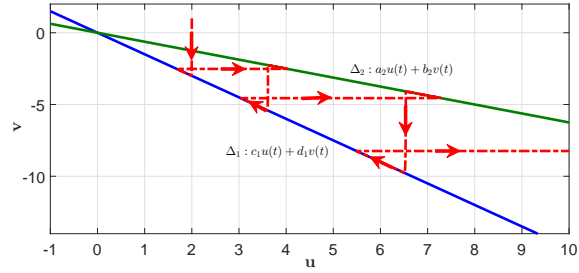}
\caption{State's trajectory in $(u,v)$- plane for \eqref{mode1}-\eqref{mode2} with $\AA_1, \AA_2$ defined by \eqref{num-eval2} and  $t_{k+1}-t_k=\tau = 0.16 \ sec$}\label{manifold5}
 \end{figure}

The two slow manifolds associated with the system are given in this case by the lines:
\[\left\{\begin{split}
&-3 u(t)-2v(t)=0\\
&-2.5 u(t)- 4 v(t)=0.
\end{split}\right.
\]
As previously we consider the initial condition $\x_0=(2,\ 1)$, $\varepsilon = 10^{-3}$,
simulating the system with an inter-events period given by $t_{k+1}-t_k=0.16\ sec,\ \forall k\in\N,$ one can observe from Fig. \ref{manifold5} that system's trajectory diverges.

 \vspace{-3 mm}
 
\begin{figure}[!h]
\centering \includegraphics[scale=0.33]{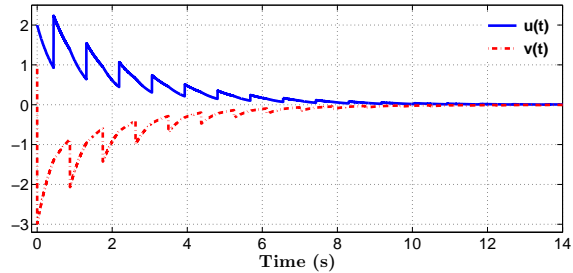}
\caption{State's trajectory for  \eqref{mode1}-\eqref{mode2} with $\AA_1, \AA_2$ defined by \eqref{num-eval2} and  $t_{k+1}-t_k=\tau = 0.406 \ sec$}\label{Traj3} \end{figure}

\begin{figure}[!h]
\centering \includegraphics[scale=0.33]{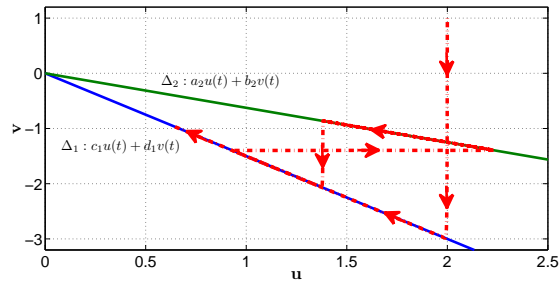}
\caption{First part of the state's trajectory in $(u,v)$- plane for \eqref{mode1}-\eqref{mode2} with $\AA_1, \AA_2$ defined by \eqref{num-eval2} and  $t_{k+1}-t_k=\tau = 0.406\ sec$}\label{manifold6}
 \end{figure}

Indeed, using Theorem \ref{theorem_final} we obtain a required dwell-time equals $0. 406\ sec$ to ensure stability. Simulating the system again with $t_{k+1}-t_k=\tau = 0.406 \ sec$ we can see in Fig. \ref{Traj3} that expected stability is obtained. Fig. \ref{manifold6} shows the first part of the trajectory illustrating its behavior.

\section{Conclusion}

Motivated by a real problem in steel production we introduced and analyzed a class of singularly perturbed switched linear systems in which the nature of the variable is mode-dependent. At switching instants, slow variables can become fast and reversely. Moreover, the state vector can loose or gain components at the switchings times. We show that the dwell-time required to ensure stability of the overall system is the sum of two terms. 
\blue{The first one essentially consists of a dwell-time ensuring stability of the reduced order system. The second term depends on the scale parameter defining the ratio between the two time-scales and goes to zero when the parameter goes to zero.
Our results complement existing results on stability analysis of singularly perturbed linear systems by showing the correlation between the values of the stabilizing dwell-time and of the scale parameter.} A low-dimension numerical example illustrates our results.

\blue{\appendix
{\it Proof of Proposition \ref{Proposition1}:} By computing the time derivative of $V$ along the trajectories of \eqref{equation_simple}, one has 
\begin{equation*}
\begin{split}
\dot{V} = &\; 2x^{\top} Q_s \dot{x}+ 2y^{\top} Q_f \dot{y} 
= 2 x^{\top} Q_s A_0 x + \frac{2}{\varepsilon} y^{\top} Q_f A_{22} y \\
& \;  +2 x^{\top} Q_s B_1 y  + 2 y^{\top} Q_f B_2x +  2 y^{\top} Q_f B_3 y\\
\le &\; 
 -2 \lambda_s x^{\top} Q_s x - \frac{2\lambda_f}{\varepsilon}y^{\top}Q_f y \\ 
 &\;  + 2 (b_1 +b_2) \sqrt{x^{\top} Q_s x}\sqrt{y^{\top} Q_f y} +2 b_3 y^{\top} Q_f y\\
\end{split}
\end{equation*}
Then, it follows that
\begin{equation*}
\dot{V}
\leq -\left(\frac{2\lambda_f}{\varepsilon}-2b_3- \frac{(b_1+b_2)^{2}}{2\lambda_s}\right) y^{\top} Q_f y.\\
\end{equation*}
Then, for all $\varepsilon \in(0,\varepsilon_1]$ , $\dot V \le 0$. Since $V$ is also positive definite and radially unbounded, it is a Lyapunov function for system \eqref{equation_simple}.}

\blue{{{\it Proof of Proposition \ref{Proposition2} :}} Computing the time derivative of $W_f$ gives
\begin{equation*}
\begin{split}
\dot{W}_f&=\frac{2 y^{\top}Q_f \dot{y}}{2\sqrt{y^{\top} Q_f y}}\\
&\leq \frac{-\frac{\lambda_f}{\varepsilon} y^{\top}Q_f y+y^{\top}Q_f (B_2 x+B_3 y)}{\sqrt{y^{\top} Q_f y}}\\
&\leq -\frac{\lambda_f}{\varepsilon} W_f +b_2 W_s+b_3 W_f\\
&\leq -\frac{\lambda_f}{\varepsilon} W_f +\sqrt{b_2^2+b_3^2}\sqrt{W_s^2+W_f^2} \\
&\leq -\frac{\lambda_f}{\varepsilon} W_f +\sqrt{b_2^2+b_3^2}\sqrt{V}. \\
\end{split}
\end{equation*}
From Proposition \ref{Proposition1}, it follows that for all $t\ge 0$,
\begin{equation*}
\dot{W}_f(t)\leq  -\frac{\lambda_f}{\varepsilon} W_f(t) + \sqrt{b_2^2+b_3^2}\sqrt{V(0)}. 
\end{equation*}
Then, we have
\begin{equation*}
\begin{split}
W_f(t) \leq \; & W_f(0) \e^{-\frac{\lambda_f}{\varepsilon}t}+   \sqrt{b_2^2+b_3^2} \sqrt{V(0)} \int_0^{t} e^{-\frac{\lambda_f}{\varepsilon}(t-s)} ds \\
 \leq \; & W_f(0) \e^{-\frac{\lambda_f}{\varepsilon}t}+  \varepsilon \frac{ \sqrt{b_2^2+b_3^2}}{\lambda_f} \sqrt{V(0)}  \big( 1 - \e^{-\frac{\lambda_f}{\varepsilon}t} \big)\\
 \leq \; & W_f(0) \e^{-\frac{\lambda_f}{\varepsilon}t}+  \varepsilon \frac{ \sqrt{b_2^2+b_3^2}}{\lambda_f} \sqrt{V(0)}.  
\end{split}
\end{equation*}}

\blue{{\it Proof of Proposition \ref{Proposition3} :} Computing the time derivative of $W_s$ gives
\begin{equation*}
\begin{split}
\dot{W}_s=\; & \frac{2x^{\top}Q_s \dot{x}}{2\sqrt{x^{\top} Q_s x}} 
\le \frac{- \lambda_s x^{\top}Q_s x+ x^{\top} Q_s B_1 y}{\sqrt{x^{\top} Q_s x}}\\
\leq\; &  -\lambda_s W_s + b_1 W_f.\\
\end{split}
\end{equation*}
Using Proposition~\ref{Proposition2}, one gets:
\begin{equation*}
\begin{split}
\dot{W}_s(t)
\leq\; & -\lambda_s W_s(t)  +b_1 W_f(0) \e^{\frac{-\lambda_f}{\varepsilon} t} + \varepsilon  b_1 \beta_1 \sqrt{V(0)}.
\end{split}
\end{equation*}
Then, we have:
\begin{equation*}
\begin{split}
W_s(t) \leq\;& \e^{-\lambda_s t}W_s(0)+ b_1 W_f(0) \int_0^{t} \e^{- \frac{\lambda_f}{\varepsilon} s} \e^{-\lambda_s (t-s)} ds\\ & +\varepsilon b_1 \beta_1 \sqrt{V(0)} \int_0^{t}  \e^{-\lambda_s (t-s)} ds\\
\leq\; & \e^{-\lambda_s t}W_s(0)+  \frac{b_1}{\frac{\lambda_f}{\varepsilon}-\lambda_s} W_f(0) \Big( \e^{-\lambda_s t}-\e^{-\frac{\lambda_f}{\varepsilon}t}\Big)\\ 
& + \frac{\varepsilon b_1 \beta_1 }{\lambda_s}\sqrt{V(0)} \Big( 1- \e^{-\lambda_s t} \Big).
\end{split}
\end{equation*}
Then, $\varepsilon \le \varepsilon_2 < \frac{\lambda_f}{\lambda_s}$ gives
\begin{equation*}
\begin{split}
W_s(t) 
\leq\; & \e^{-\lambda_s t}W_s(0)+  \frac{b_1\varepsilon}{\lambda_f-\varepsilon_2\lambda_s}W_f(0) +\frac{\varepsilon b_1 \beta_1}{\lambda_s} \sqrt{V(0)}.  \\
\end{split}
\end{equation*}
}

\blue{{\it Proof of Theorem \ref{theorem_final}:} Let us remark that
\begin{equation*}
\Gamma M_{\uptau^*} = 
 \left(
\begin{array}{cc}
\gamma_{11}e^{-\lambda_s \uptau^*}+\varepsilon \delta_{1} & \gamma_{12}e^{-\frac{\lambda_f}{\varepsilon}\uptau^*} +\varepsilon \delta_2 \\
\gamma_{21}e^{-\lambda_s \uptau^*}+\varepsilon \delta_{3} & \gamma_{22}e^{-\frac{\lambda_f}{\varepsilon}\uptau^*} +\varepsilon \delta_4 
\end{array}
\right)
\end{equation*}
where
\begin{equation}
\label{eq:delta}
\begin{split}
 \delta_{1} =\;& \gamma_{11} \beta_3+\gamma_{12}\beta_1,\; \delta_2=\gamma_{11}(\beta_2+\beta_3)+\gamma_{12}\beta_1,\\
 \delta_3= \;& \gamma_{21} \beta_3+\gamma_{22}\beta_1,\; \delta_4=\gamma_{21}(\beta_2+\beta_3)+\gamma_{22}\beta_1.
\end{split}
\end{equation}
Moreover, the positive matrix $\Gamma M_{\uptau^*}$ is Schur if and only if there exists $p\in \R_+^2$, such that $\big(\Gamma M_{\uptau^*}\big)^\top p < p$ (see e.g.  \cite{rantzer2011distributed}). Let us look for $p$ under the form $(1,a\varepsilon)^\top$ with $a>\delta_2$. Then, $\big(\Gamma M_{\uptau^*}\big)^\top p < p$ is equivalent to
\begin{equation}
\label{eq:dwell}
\left\{
\begin{array}{lll}
\gamma_{11} e^{-\lambda_s \uptau^*} +\varepsilon \delta_1 +a \varepsilon \gamma_{21}e^{-\lambda_s \uptau^*}+a \varepsilon^2 \delta_{3} &<& 1 \\
\gamma_{12}e^{-\frac{\lambda_f}{\varepsilon}\uptau^*} +\varepsilon \delta_2 + a \varepsilon \gamma_{22}e^{-\frac{\lambda_f}{\varepsilon}\uptau^*} +a\varepsilon^2 \delta_4 &<& a\varepsilon.
\end{array}\right.
\end{equation}
The first inequality of \eqref{eq:dwell} is equivalent to
\begin{equation*}
 \uptau^* > \frac{-1}{\lambda_s}\ln\Big(\frac{1-\varepsilon \delta_1 -a \varepsilon^2 \delta_{3}}{\gamma_{11} + a \varepsilon \gamma_{21}}\Big)=\frac{\ln(\gamma_{11})}{\lambda_s}+\eta_1(\varepsilon). 
\end{equation*}
where
\begin{equation}
\label{eq:eta1}
\eta_1(\varepsilon) = \frac{1}{\lambda_s}\Big(\ln(1+\frac{a \varepsilon \gamma_{21}}{\gamma_{11}})-\ln(1-\varepsilon \delta_1 -a \varepsilon^2 \delta_{3})\Big).
\end{equation}
It is easy to show that $\eta_1(\varepsilon)=\mathcal O (\varepsilon)$. Moreover, let us remark that $\eta_1(\varepsilon)$ is only defined if 
$1-\varepsilon \delta_1 -a \varepsilon^2 \delta_{3}>0$, that is if $\varepsilon < \varepsilon_3$ where
\begin{equation}
\label{eq:epsilon3}
\varepsilon_3= \frac{-\delta_1 +\sqrt{\delta_1^2+4a\delta_3}}{2a\delta_3}.
\end{equation}
The second inequality of \eqref{eq:dwell} is equivalent to
\begin{equation*}
\begin{split}
& \uptau^* > \frac{-\varepsilon}{\lambda_f}\ln\Big(\frac{a\varepsilon-\varepsilon \delta_2 -a \varepsilon^2 \delta_{4}}{\gamma_{12} + a \varepsilon \gamma_{22}}\Big) \\
\iff & \uptau^* > \frac{\varepsilon}{\lambda_f}\Big(\ln\Big(\frac{\gamma_{12} + a \varepsilon \gamma_{22}}{a- \delta_2 -a \varepsilon \delta_{4}}\Big) -\ln(\varepsilon)  \Big).
\end{split}
\end{equation*}
As $\uptau^* > \frac{\ln(\gamma_{11})}{\lambda_s}+\eta_1(\varepsilon)\ge  \frac{\ln(\gamma_{11})}{\lambda_s}$, then the previous inequality holds if
\begin{equation}
\label{eq:cs}
 \frac{\ln(\gamma_{11})}{\lambda_s} > \frac{\varepsilon}{\lambda_f}\Big(\ln\Big(\frac{\gamma_{12} + a \varepsilon \gamma_{22}}{a- \delta_2 -a \varepsilon \delta_{4}}\Big) -\ln(\varepsilon)  \Big).
\end{equation}
By remarking that the right-hand side of the inequality goes to $0$ when $\varepsilon$ goes to $0$, one concludes that exists $\varepsilon_4>0$ such that for all $\varepsilon \in (0,\varepsilon_4)$, \eqref{eq:cs} holds. Then, the theorem is proved by setting $\varepsilon_1^*=\min(\varepsilon_2,\varepsilon_3,\varepsilon_4)$.}

\blue{{\it Proof of Theorem \ref{theorem_final2}:} Similarly to the proof of Theorem~\ref{theorem_final}, it is sufficient to show that (\ref{eq:dwell}) holds. Since $\gamma_{11}=1$, the first inequality holds if and only if
$\uptau^* > \eta_1(\varepsilon)$.
The second inequality holds if and only if 
\begin{equation*}
\uptau^* > \frac{-\varepsilon}{\lambda_f}\ln(\varepsilon)+\eta_2(\varepsilon),
\end{equation*} 
where
\begin{equation}
\label{eq:eta2}
\eta_2(\varepsilon) =  \frac{\varepsilon}{\lambda_f} \ln\Big(\frac{\gamma_{12} + a \varepsilon \gamma_{22}}{a- \delta_2 -a \varepsilon \delta_{4}}\Big).
\end{equation}
It is easy to show that $\eta_2(\varepsilon) = \mathcal O(\varepsilon)$. Moreover let us remark that $\eta_2(\varepsilon)$ is only defined if $a-\delta_2-a\varepsilon \delta_{4}>0$, that is if $\varepsilon <\varepsilon_5$ with 
\begin{equation}
\label{eq:epsilon5}
\varepsilon_5= \frac{a-\delta_2}{a \delta_4}.
\end{equation}
Moreover, since $\eta_1(\varepsilon)=\mathcal O (\varepsilon)$, there exists $\varepsilon_6>0$ such that for all $\varepsilon \in (0,\varepsilon_6)$,
$\eta_1(\varepsilon) < \frac{-1}{\lambda_f}\varepsilon\ln(\varepsilon)$.
The theorem is proved by setting $\varepsilon_2^*=\min(\varepsilon_2,\varepsilon_5,\varepsilon_6)$.}

\blue{{\it Proof of Theorem  \ref{theorem_final3}:} Similarly to the proof of Theorem~\ref{theorem_final}, it is sufficient to show that (\ref{eq:dwell}) holds. Since $\gamma_{11}=1$, the first inequality holds if and only if
$\uptau^* > \eta_1(\varepsilon)$.
Since $\gamma_{12}=0$, the second inequality holds if and only if 
\begin{equation*}
\uptau^* > \frac{-\varepsilon}{\lambda_f}\ln\Big(\frac{a- \delta_2 -a \varepsilon \delta_{4}}{ a \gamma_{22}}\Big). 
\end{equation*} 
Let 
\begin{equation}
\label{eq:eta3}
\eta_3(\varepsilon)=\max\Big(\eta_1(\varepsilon), \frac{-\varepsilon}{\lambda_f}\ln\Big(\frac{a- \delta_2 -a \varepsilon \delta_{4}}{ a \gamma_{22}}\Big)\Big).
\end{equation}
Then, it is easy to show that $\eta_3(\varepsilon)=\mathcal O(\varepsilon)$ and is well defined for $\varepsilon < \min (\varepsilon_3,\varepsilon_5)$.
The theorem is proved by setting $\varepsilon_3^*=\min(\varepsilon_2,\varepsilon_3,\varepsilon_5)$.
}

\blue{{\it Proof of Theorem \ref{theorem_final4}:} The positive matrix $\Gamma M_{\uptau^*}$ is Schur if and only if there exists $p\in \R_+^2$, such that $\big(\Gamma M_{\uptau^*}\big)^\top p < p$ (see e.g.  \cite{rantzer2011distributed}). Let us look for $p$ under the form $(1,a)^\top$ with $a>0$. Then, $\big(\Gamma M_{\uptau^*}\big)^\top p < p$ is equivalent to
\begin{equation*}
\left\{
\begin{array}{lll}
\gamma_{11} e^{-\lambda_s \uptau^*} +\varepsilon \delta_1 +a  \gamma_{21}e^{-\lambda_s \uptau^*}+a \varepsilon \delta_{3} &<& 1 \\
\gamma_{12}e^{-\frac{\lambda_f}{\varepsilon}\uptau^*} +\varepsilon \delta_2 + a  \gamma_{22}e^{-\frac{\lambda_f}{\varepsilon}\uptau^*} +a\varepsilon \delta_4 &<& a
\end{array}\right.
\end{equation*}
which is also equivalent to
\begin{equation}
\label{eq:dwell3}
\left\{
\begin{array}{lll}
\uptau^* & > & \frac{1}{\lambda_s} \ln\Big( \frac{\gamma_{11}+a\gamma_{21}}{1-a\varepsilon  \delta_3 - \varepsilon \delta_1}\Big) \\
\uptau^* & > & \frac{\varepsilon}{\lambda_f} \ln\Big( \frac{\gamma_{12}+a\gamma_{22}}{a-\varepsilon  \delta_2 - a\varepsilon \delta_4}\Big). 
\end{array}\right.
\end{equation}
Since $\gamma_{11}<1$ it is possible to choose $a>0$ such that $\gamma_{11}+a \gamma_{21} <1$, it follows that the first inequality holds for any $\uptau^*\ge 0$ and 
for all $\varepsilon \in (0,\varepsilon_6)$ with
\begin{equation}
\label{eq:epsilon6}
\varepsilon_6 = \frac{1-\gamma_{11}-a\gamma_{21}}{a\delta_3+\delta_1}.
\end{equation}
Then the second inequality is equivalent to $\uptau^*>\eta_4(\varepsilon)$ where 
\begin{equation}
\label{eq:eta4}
\eta_4(\varepsilon) = \frac{\varepsilon}{\lambda_f} \ln\Big( \frac{\gamma_{12}+a\gamma_{22}}{a-\varepsilon  \delta_2 - a\varepsilon \delta_4}\Big) .
\end{equation}
It is easy to show that $\eta_4(\varepsilon) = \mathcal O(\varepsilon)$ and is well defined for $\varepsilon < \varepsilon_7$ given by
\begin{equation}
\label{eq:epsilon7}
\varepsilon_7=\frac{a}{a\delta_3+\delta_1}.
\end{equation}
The theorem is proved by setting $\varepsilon_4^*=\min(\varepsilon_2,\varepsilon_6,\varepsilon_7)$.
}

\blue{{\it Proof of Theorem \ref{theorem_final5}:} Similar to the proof of Theorem~\ref{theorem_final4}, it is sufficient to show that (\ref{eq:dwell3}) holds. Since  $\gamma_{11}< 1$, $\gamma_{22}<1$ and $\frac{\gamma_{12}\gamma_{21}}{(1-\gamma_{11})(1-\gamma_{22})}<1$, it is possible to choose $a>0$ such that 
$\frac{\gamma_{12}}{1-\gamma_{22}} < a < \frac{1-\gamma_{11}}{\gamma_{21}}$. It follows  that the first inequality holds for any $\uptau^*\ge 0$ and 
for all $\varepsilon \in (0,\varepsilon_6)$. As for the second inequality, it holds for any $\uptau^*\ge 0$ and 
for all $\varepsilon \in (0,\varepsilon_8)$ with 
\begin{equation}
\label{eq:epsilon8}
\varepsilon_8 = \frac{a-\gamma_{12}-a\gamma_{22}}{\delta_2+a\delta_4}.
\end{equation}
The theorem is proved by setting $\varepsilon_5^*=\min(\varepsilon_2,\varepsilon_6,\varepsilon_8)$.}

\bibliographystyle{IEEEtran}

\end{document}